\documentclass{aa}
\usepackage{graphicx}
\usepackage{xcolor}
\usepackage{txfonts}
\usepackage{booktabs}
\usepackage{multirow}
\usepackage{makecell}
\usepackage{nicefrac}
\usepackage{siunitx}
\usepackage{longtable}
\usepackage{subfig}
\usepackage[colorlinks=true,linkcolor=blue,allcolors=blue]{hyperref}

\newcommand{\lya}{{\text{Ly}\ensuremath{\alpha}}}
\newcommand{\Halpha}{{\text{H}\ensuremath{\alpha}}}
\newcommand{\Hbeta}{{\text{H}\ensuremath{\beta}}}
\newcommand{\Hgamma}{{\text{H}\ensuremath{\gamma}}}
\newcommand{\oiii}{{\text{[\ion{O}{iii}]}}}

\newcommand{\feii}{{\text{\ion{Fe}{ii}}}}

\newcommand{\Hden}{\ensuremath{n_{\rm H}}}
\newcommand{\NH}{\ensuremath{N_{\rm H}}}

\newcommand{\eden}{\ensuremath{n_{\rm e}}}

\newcommand{\Te}{\ensuremath{T_{\rm e}}}
\newcommand{\logten}{\ensuremath{\log_{10}}}
\newcommand{\vturb}{\ensuremath{v_{\rm turb}}}
\newcommand{\logU}{\ensuremath{\logten U}}

\defcitealias{naidu2024}{Naidu \& Matthee et al. 2024}

\let\unit\si

\begin{document}

   \title{The warm outer layer of a little red dot as the source of [\feii{}] and collisional Balmer lines with scattering wings}
   
   \titlerunning{The origins of emission lines from a luminous little red dot}

   \author{
   Alberto~Torralba\inst{\ref{inst:ista}}\thanks{alberto.torralba@ista.ac.at}
   \and Jorryt~Matthee\inst{\ref{inst:ista}}
   \and Gabriele~Pezzulli\inst{\ref{inst:kapteyn}}
   \and Rohan~P.~Naidu\inst{\ref{inst:MIT_kavli}}
   \and Yuzo~Ishikawa\inst{\ref{inst:MIT_kavli}}
   \and Gabriel~B.~Brammer\inst{\ref{inst:dawn}, \ref{inst:bohr}}
   \and Seok-Jun~Chang\inst{\ref{inst:MPIA_garching}}
   \and John~Chisholm\inst{\ref{inst:texas_austin}}
   \and Anna~de~Graaff\inst{\ref{inst:mpia}}
   \and Francesco~D'Eugenio\inst{\ref{inst:kavli_cambr}, \ref{inst:cavendish_cambr}}
   \and Claudia~Di~Cesare\inst{\ref{inst:ista}}
   \and Anna-Christina~Eilers\inst{\ref{inst:MIT_kavli}}
   \and Jenny~E.~Greene\inst{\ref{inst:princeton}}
   \and Max~Gronke\inst{\ref{inst:astro_heidelberg}, \ref{inst:MPIA_garching}}
   \and Edoardo~Iani\inst{\ref{inst:ista}}
   \and Vasily~Kokorev\inst{\ref{inst:texas_austin}}
   \and Gauri~Kotiwale\inst{\ref{inst:ista}}
   \and Ivan~Kramarenko\inst{\ref{inst:ista}}
   \and Yilun~Ma\inst{\ref{inst:princeton}}
   \and Sara~Mascia\inst{\ref{inst:ista}}
   \and Benjamín~Navarrete\inst{\ref{inst:ista}}
   \and Erica~Nelson\inst{\ref{inst:boulder}}
   \and Pascal~Oesch\inst{\ref{inst:unige}, \ref{inst:dawn}}
   \and Robert~A.~Simcoe\inst{\ref{inst:MIT_phys_dpt}, \ref{inst:MIT_kavli}}
   \and Stijn~Wuyts\inst{\ref{inst:bath}}
   }

   \institute{
    Institute of Science and Technology Austria (ISTA), Am Campus 1, 3400 Klosterneuburg, Austria.\label{inst:ista}
    \and Kapteyn Astronomical Institute, University of Groningen, Landleven 12, NL-9747 AD Groningen, the Netherlands.\label{inst:kapteyn}
    \and MIT Kavli Institute for Astrophysics and Space Research, Massachusetts Institute of Technology, Cambridge, MA 02139, USA.\label{inst:MIT_kavli}
    \and Cosmic Dawn Center (DAWN), Niels Bohr Institute, University of Copenhagen, Jagtvej 128, K\o benhavn N, DK-2200, Denmark\label{inst:dawn}
    \and Niels Bohr Institute, University of Copenhagen, Jagtvej 128, 2200 Copenhagen N, Denmark\label{inst:bohr}
    \and Max Planck Institute for Astrophysics, Karl-Schwarzschild-Str. 1, 85748 Garching, Germany.\label{inst:MPIA_garching}
    \and Department of Astronomy, University of Texas at Austin, 2515 Speedway, Austin, Texas 78712, USA\label{inst:texas_austin}
    \and Max-Planck-Institut f\"ur Astronomie, K\"onigstuhl 17, D-69117 Heidelberg, Germany\label{inst:mpia}
    \and  Kavli Institute for Cosmology, University of Cambridge, Madingley Road, Cambridge, CB3 0HA, United Kingdom\label{inst:kavli_cambr}
    \and  Cavendish Laboratory, Astrophysics Group, University of Cambridge, 19 JJ Thomson Avenue, Cambridge, CB3 0HE, United Kingdom\label{inst:cavendish_cambr}
    \and Department of Astrophysical Sciences, Princeton University, Princeton, NJ 08544, USA.\label{inst:princeton}
    \and Astronomisches Rechen-Institut, Zentrum für Astronomie, Universität Heidelberg, Mönchhofstraße 12-14, 69120 Heidelberg, Germany\label{inst:astro_heidelberg}
    \and Department for Astrophysical and Planetary Science, University of Colorado, Boulder, CO 80309, USA\label{inst:boulder}
    \and Department of Astronomy, University of Geneva, Chemin Pegasi 51, 1290 Versoix, Switzerland\label{inst:unige}
    \and Department of Physics, Massachusetts Institute of Technology, Cambridge, MA 02139, USA\label{inst:MIT_phys_dpt}
    \and Department of Physics, University of Bath, Claverton Down, Bath BA2 7AY, UK\label{inst:bath}
    }

  \date{Accepted XXX. Received YYY; in original form ZZZ}

  \abstract{
    The population of the little red dots (LRDs) may represent a key phase of supermassive black hole (SMBH) growth. A cocoon of dense excited gas is emerging as a key component to explain the most striking properties of LRDs, such as strong Balmer breaks and Balmer absorption, as well as the weak IR emission. To dissect the structure of LRDs, we analyzed new deep JWST/NIRSpec PRISM and G395H spectra of FRESCO-GN-9771, one of the most luminous known LRDs at $z=5.5$. These spectra reveal a strong Balmer break, broad Balmer lines, and very narrow [O\,{\sc iii}] emission. We revealed a forest of optical [Fe\,{\sc ii}] lines, which we argue are emerging from a dense ($n_{\rm H}=10^{9-10}$ cm$^{-3}$) warm layer with electron temperature $T_{\rm e}\approx7000$ K. The broad wings of H$\alpha$ and H$\beta$ have an exponential profile due to electron scattering in this same layer. The high $\rm H\alpha:H\beta:H\gamma$ flux ratio of $\approx10.4:1:0.14$ is an indicator of collisional excitation and resonant scattering dominating the Balmer line emission. A narrow H$\gamma$ component, unseen in the other two Balmer lines due to outshining by the broad components, could trace the ISM of a normal host galaxy with a star formation rate of $\sim5$ M$_{\odot}$ yr$^{-1}$. The warm layer is mostly opaque to Balmer transitions, producing a characteristic  P Cygni profile in the line centers suggesting outflowing motions. This same layer is responsible for shaping the Balmer break. The  broadband spectrum can be reasonably matched by a simple photoionized slab model that dominates the $\lambda>1500$ {\AA} continuum and a low-mass ($\sim10^8$ M$_{\odot}$) galaxy that could explain the narrow [O\,{\sc iii}], with only a subdominant contribution to the UV continuum. Our findings indicate that Balmer lines are not directly tracing the gas kinematics near the SMBH and that the BH mass scale is likely much lower than virial indicators suggest.
  }


   \maketitle

\section{Introduction}\label{sec:introduction}

JWST has identified a population of faint compact broad Balmer line emitters at redshifts $z=2$--9 that are providing new insights into the formation of supermassive black holes. A significant fraction of these show blue UV slopes, but red UV to optical colors \citep[e.g.,][]{labbe2024, akins2024, kokorev2024a, barro2024} that are often due to a Balmer break \citep[e.g.,][]{setton2024,ji2025}. This combination of features makes these sources appear red in the NIRCam images, hence the name little red dots (LRDs; \citealt{matthee2024a}). 

Little red dots constitute about $\sim$1\% of the galaxy population at $z\sim5$, and they are about 100--1000 times more numerous than similarly luminous UV-selected quasars at the same redshifts \citep[e.g.,][]{harikane2023,kocevski2023,akins2024,kokorev2024a,greene2024,maiolino2024a,matthee2024a, Lin25}. LRDs have already been detected spectroscopically out to $z=9$ \citep{Taylor25z9}, and their number densities decline at lower redshifts \citep[$z\lesssim 4$; e.g.,][]{kocevski2024, ma2025_counting, inayoshi2025_decline,loiacono2025} suggesting that the physical conditions in the early Universe are more favorable to LRD formation. LRDs appear to be hosted by low-mass ($M_*\lesssim 10^9\ M_\odot$) galaxies, as indicated by spectral energy distribution (SED) fitting studies \citep[e.g.,][]{maiolino2024a,wang2024, ma2025}, clustering measurements \citep{matthee2024b,pizzati2024,carranza25}, and the faintness of their extended rest-UV counterparts (undetected in many cases; e.g., \citealt{rinaldi24,chen2025, torralba2025}). 

The mechanism that powers the emission of LRDs has been the subject of significant debate \citep[e.g.,][]{perez-gonzalez2024,akins2024,greene2024,baggen24}, but is primarily thought to be driven by accretion onto supermassive black holes (SMBH) due to their compactness and their broad hydrogen lines (FWHM~$\gtrsim 1000$~\unit{km.s^{-1}}) that are reminiscent of those of typical Type I active galactic nuclei \citep[AGN; e.g.,][]{greene2024}. However, the spectra of LRDs display several differences with quasars and other types of AGN. They generally lack strong broad high-ionization lines in the UV (such as \ion{C}{iv}, although some narrow high-ionization lines have been reported in a few of them; \citealt{treiber2024, tang2025}), they lack X-ray emission at the level expected for their H$\alpha$ emission \citep[e.g.,][]{yue2024, ananna2024}, and they do not show any detectable radio emission \citep[][]{latif2025, perger2025, gloudemans2025, mazzolari2024}. Moreover, a dust-reddened AGN is unlikely responsible for the red UV to optical colors, as indicated by the absence of far-infrared emission \citep[e.g.,][]{williams2024, leung2024, Setton2025,Xiao25}. These differences with typical AGN raise caveats when trying to estimate their intrinsic AGN properties (BH mass, bolometric luminosity) using standard scaling relations \citep[e.g.,][]{sacchi2025, degraaff2025, rusakov2025, naidu2025}.

Deep spectra of luminous LRDs are also revealing fainter emission features that hold  important clues to  their nature. In particular, \feii{} emission is seen in the rest-frame optical range \citep[e.g.,][]{lin2025, ji2025_LoL}, and it appears unusually strong in the rest-frame UV \citep{labbe2024, tripodi_deep_2025}. \feii{} emission is rare in regular galaxies, but is commonly seen in AGN, and it is associated with  dense clouds with high column density within broad-line regions  (BLRs) \citep{netzer1974, oke1979, veron2002, veron-cetty2004}.
\feii{} emission is characteristic of the  BLR of classical AGN \citep{baldwin2004, ferland2009, marinello2016, gaskell2022}, where it originates within the dust-sublimation radius (as otherwise iron atoms would be depleted onto dust grains). Some AGN also present narrower forbidden [\feii{}] lines in their optical spectra, such as Seyfert 1 galaxies \citep{netzer1974, oke1979, veron2002, veron-cetty2004}, and rare cases of quasars with Balmer absorptions and strong \ion{He}{i} lines \citep[e.g.,][]{wang2008}. Interestingly, the forbidden [\feii{}] emission is also found in objects with dense gas envelopes such as Type IIn supernovae \citep[e.g.,][]{groeningsson2007,dessart2009}, supernova remnants \citep[e.g.,][]{koo2014, lee2019, alistecastillo2025}, the envelopes of luminous blue variable stars \citep[e.g.,][]{hillier2001, peng2025} and Be stars \citep[e.g.,][]{Arias06}.

The presence of a very dense gas ($\Hden\gtrsim 10^8$~\unit{cm^{-3}}) appears to play an important role in the spectra of LRDs. The presence of a layer of gas with a high neutral hydrogen column density in the 2s state could explain the strong observed Balmer breaks \citep{inayoshi2025, ji2025}, especially those stronger than any stellar population can produce \citep{naidu2025, degraaff2025}. The same dense gas would provide an explanation for the unusual absorption detected in Balmer lines \citep[e.g.,][]{matthee2024a, juodzbalis2024, deugenio2025,kocevski2024}, it would mitigate the need of significant dust attenuation \citep[e.g.,][]{naidu2025}, therefore providing an explanation for the far-infrared non-detections, and possibly provide an explanation to the absence of X-rays due to Compton-thick absorption \citep[e.g.,][]{juodzbalis2025, maiolino2025},  although a simple reason for the lack of X-rays could be that this part of the spectrum is intrinsically weak \citep{BWang25}.

Simple Cloudy\footnote{\url{https://gitlab.nublado.org/cloudy/cloudy/-/wikis/home}} models have been remarkably successful in explaining various observed characteristics of LRD spectra \citep[e.g.,][]{ji2025, naidu2025, degraaff2025}. These models put a slab of dense gas struck by a source (usually a standard AGN spectrum). The slab is optically thick to UV, and H-ionizing radiation in general, so it reprocesses the internal energy source and reradiates it into the optical with a shape roughly similar to a blackbody. Theoretically, configurations of a high covering fraction of dense gas around a growing massive black hole are being investigated in the context of spherical accretion \citep[e.g.,][]{liu2025} and quasi stars and envelopes \citep[e.g.,][]{kido2025, begelman2025}, but we note that in these models the intrinsic AGN spectrum also differs from a standard one.

Although dense gaseous envelopes appear to be an important constituent of LRDs, many questions remain regarding the properties of the envelopes (i.e., density, temperature, gas dynamics; \citealt{nandal2025, begelman2025, ji2025_LoL}),  whether the gas fully covers the central source \citep[e.g.,][]{naidu2025, rusakov2025, torralba2025, lin2025}, and the way in which these envelopes impact our measurements of the BH mass (i.e., radiative transfer effects impacting the Balmer lines; e.g., \citealt{rusakov2025,Chang25}).

The physical conditions of such envelopes can be probed with sensitive spectroscopy. We recently observed five LRDs discovered using NIRCam grism spectroscopy from the FRESCO survey \citep[PID 1895; PI Oesch;][]{oesch2023} with the JWST/NIRSpec IFU mode. Here we focus on FRESCO-GN-9771 (hereafter GN-9771), an LRD at $z=5.535$ that has the most luminous H$\alpha$ line among the sample of \cite{matthee2024a} with $L_\Halpha = 4.5 \times 10^{43}$~\unit{erg.s^{-1}} and magnitude of F444W~$=23.0$, making it one of the most luminous LRDs known, closely following A2744-45924 presented in \citet{labbe2024}. GN-9771 is unresolved in NIRCam imaging data, and the grism spectrum was one of the first to show H$\alpha$ absorption. One of the key goals of our program is to obtain sensitive spectroscopy at the highest resolution possible for JWST covering both the H$\alpha$ and the H$\beta$ lines. Additionally, Prism spectroscopy enables the full characterization of the rest-frame UV to optical spectrum.

This paper is structured as follows. In Sect.~\ref{sec:data} we present the JWST/NIRSpec IFU data used for the analysis of the object GN-9771. In Sect.~\ref{sec:results} we present the observed spectral properties of GN-9771. In Sect.~\ref{sec:phot_models} we describe the photoionization models used to characterize the observed spectral features. In Sect.~\ref{sec:interpretation} we interpret the observational results in the context of our photoionization models, and we discuss their implications for the interpretation in the full picture of the LRD population in Sect.~\ref{sec:implications}. Finally, in Sect.~\ref{sec:summary} we summarize the contents of this paper.

Throughout this work we use a $\Lambda$CDM cosmology as described by \citet{collaboration2020}, with $\Omega_\Lambda=0.69$, $\Omega_\text{M}=0.31$, and $H_0=67.7$ km\,s$^{-1}$\,Mpc$^{-1}$. All photometric magnitudes are given in the AB system \citep{oke1983}.


\section{Data} \label{sec:data}

 We used JWST/NIRSpec IFU spectroscopy from the Cycle 4 program PID 5664 (PI Matthee). GN-9771 was observed for 9.5 hours on Jan 20, 2025. The IFU mode was used with the PRISM ($R\approx 100$) and high-resolution grating ($R\approx 3000$) G395H dispersers. In both modes, we used eight dithers with a medium cycling pattern that offers a compromise between a large enough step size to mitigate microshutter failures, while ensuring good sub-pixel sampling and a large FoV $\sim2.5\arcsec\times2.5\arcsec$ that has the full exposure time, similar to earlier successful IFU observations of faint AGN at $z\sim5.5$ \citep{Uebler23}.

\begin{figure*}
    \centering
    \includegraphics[width=0.8\linewidth]{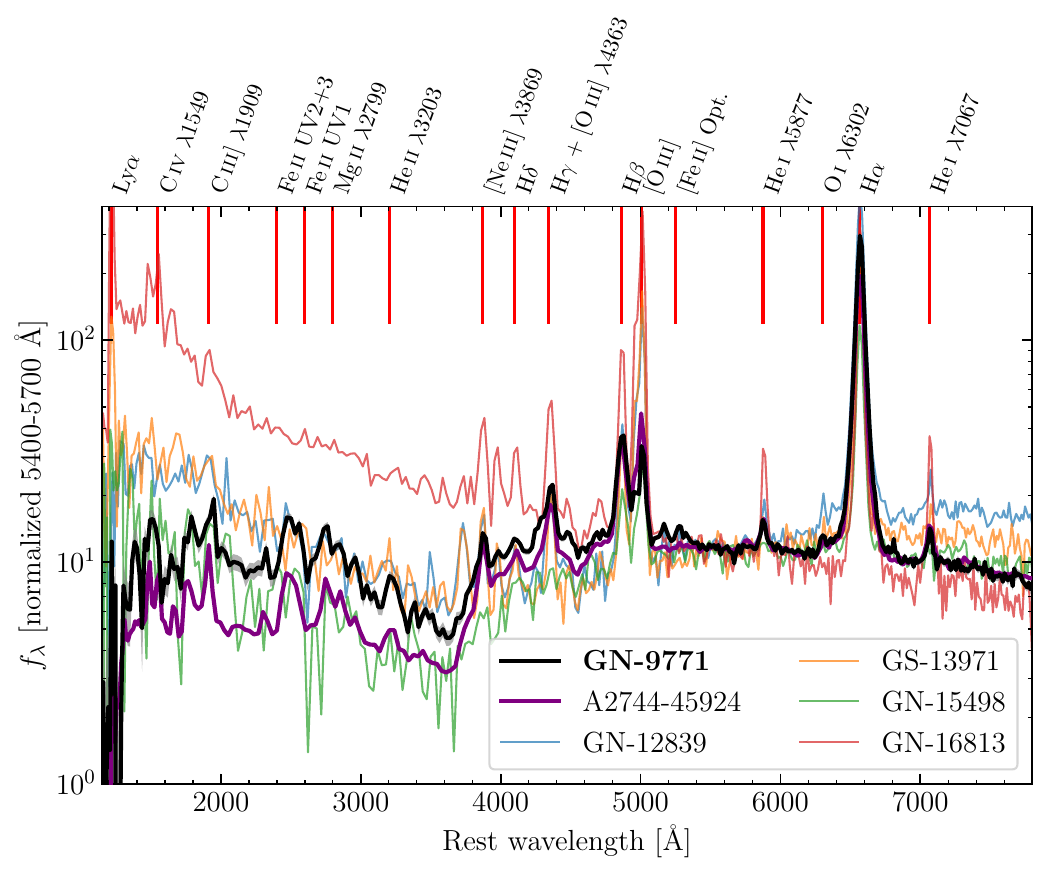}
    \caption{{\bf Broadband spectrum of GN-9771 in comparison to other LRDs.} We have highlighted the main emission lines on the top axis. We make a comparison with the PRISM spectrum of the other LRDs in our IFU program (Matthee et al. in prep.) and of A2744-45924 \citep{labbe2024}, after normalizing their flux to match GN-9771 in the 5400--5700~\AA{} band. The spectrum of GN-9771 is very similar to the spectrum of A2744-45924, with only a slightly weaker Balmer break in the case of GN-9771.}
    \label{fig:prism_spectrum}
\end{figure*}

 The PRISM observations were used to characterize the rest-frame UV-to-optical spectrum and total 6.5 ks of exposure time. The G395H/F290LP grating observations were designed to obtain sensitive high-resolution spectra for the H$\beta$, \oiii{} and H$\alpha$ emission lines and total 18.2 ks of exposure time. An optimally extracted 1D spectrum based on the PRISM observations of GN-9771 is shown in Fig.~\ref{fig:prism_spectrum}, where we also plot the rest of the sample observed by our NIRSpec program (Matthee et al. in prep.), and A2744-45924 \citep{labbe2024}, which has a remarkably similar spectrum.

We refer to Ishikawa et al. (in prep.) for a detailed description of the NIRSpec IFU data reduction. The data reduction was completed using the STScI JWST pipeline\footnote{\url{https://github.com/spacetelescope/jwst}} version 1.17.1 with Calibration Reference Data System version 12.0.9 (\texttt{jwst\_1299.pmap}). The first stage, \texttt{Detector1Pipeline}, performs standard infrared detector reductions. We correct for the $1/f$ noise \citep{Schlawin2020} on each detector by using a running mean algorithm.  We also use \texttt{snowblind}\footnote{\url{https://github.com/mpi-astronomy/snowblind}} to remove noise from snowball effects and cosmic rays. Then, these rate files are processed by \texttt{Spec2Pipeline} that produces calibrated spectra data, assigns the world coordinate system, and extracts the 2D spectra to build a 3D datacube using \texttt{drizzle}. Finally, we apply a sigma clip routine to mask pixels with extreme outliers and use the Photutils \texttt{reproject} method \citep{Vayner2023} to align and combine the 8 dither exposures into a single datacube with a spatial resolution of $0.05\arcsec$ per spaxel. Since no dedicated background exposures were taken, we perform aperture background subtraction from the datacubes. Both the PRISM and G395H/F290LP exposures showed non-uniform background. We detected a bright, narrow stripe that appears as an excess flux above the astrophysical background and stretches across the detector. To correct the background, we modeled the overall background (non-stripe regions) and the superimposed stripe artifact separately. Within these strips, the background was assumed to be uniform in each wavelength layer. We derived this uniform value from the running median value of empty-sky pixels, where we used a kernel of 10 wavelength layers. Empty-sky pixels were identified based on an agressive source detection in the white-light image of the PRISM and G395H data, respectively. Both components were subtracted to produce the final background-subtracted datacube. In late stages of finalizing the paper, we also reduced the data with msaexp (version 0.9.9; Brammer et al. in prep.). This reduction generally is hardly distinguishable from our standard reduction. However, the msaexp reduction yields slightly better performance in the bluest part of the G395H data, in particular because it slightly extends the wavelength coverage from 2.87 $\mu$m to $2.80 \mu$m which the crucial coverage of H$\gamma$ and \oiii{} $\lambda$4364. We have verified the flux calibration of this added range thanks to the oerlap with the available PRISM spectrum.

We find that GN-9771 appears spatially unresolved across the full wavelength coverage of our PRISM data, which is consistent with the NIRCam imaging data \citep{matthee2024a}. We also do not identify any neighbouring galaxies within the IFU field of view. Our measurements are therefore based on a 1D spectrum that is optimally extracted based on the light profile in a collapsed H$\alpha$ pseudo-narrowband image that we for simplicity derived by fitting a 2D Gaussian. Our results remain unchanged when extracting the spectrum over a simple compact circular aperture.

\section{Results: The optical emission lines of GN-9711} \label{sec:results}
In this section we provide an overview of the spectral features detected in our new deep data on GN-9771. 

\subsection{Balmer lines and {\rm \oiii{}} emission} \label{sec:balmer_O3}

\begin{figure}
    \centering
    \includegraphics[width=\linewidth]{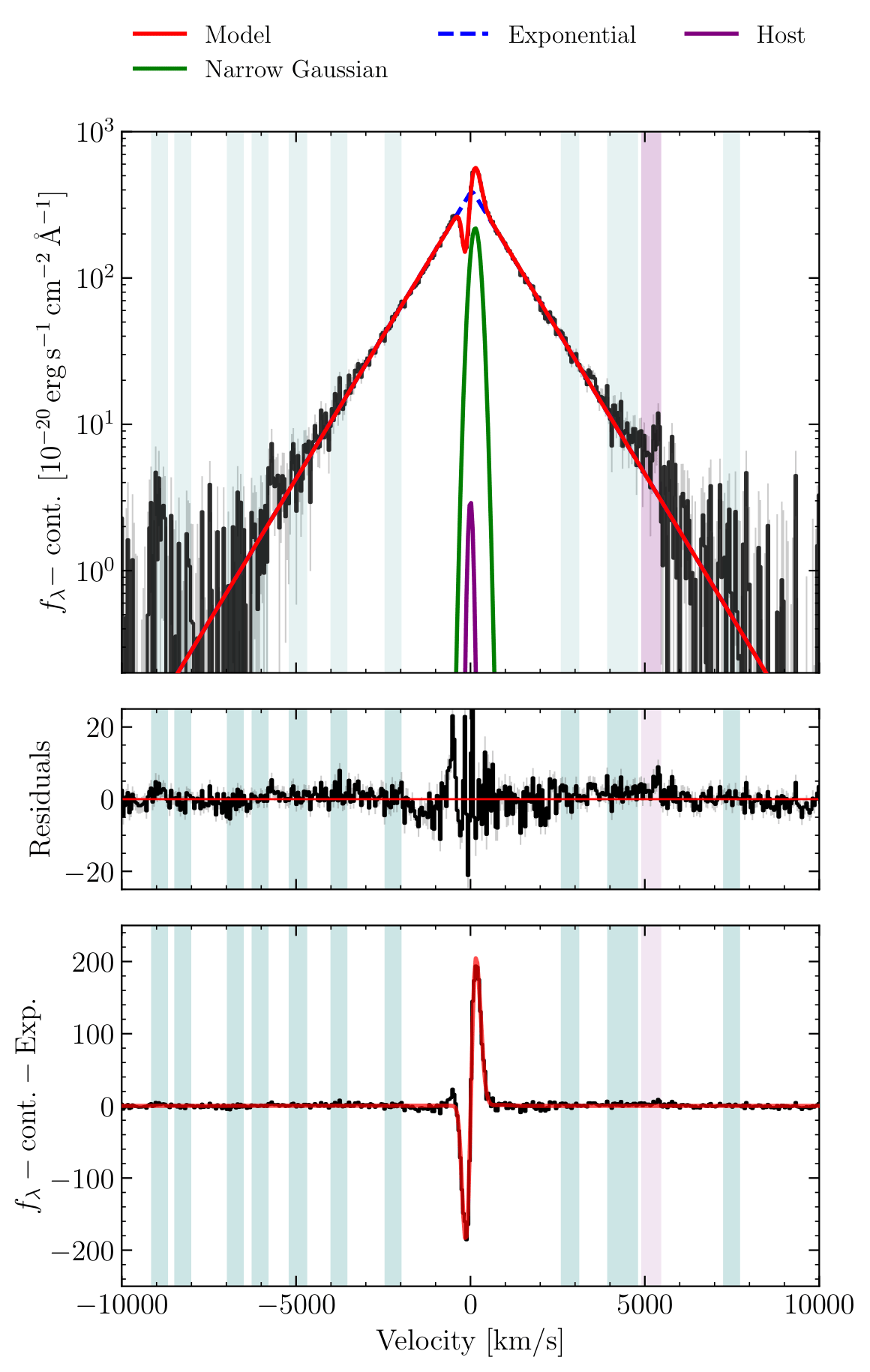}
    \caption{{\bf Triangular P Cygni H$\pmb\alpha$ spectrum of GN-9771.} The continuum-subtracted H$\alpha$ spectrum based on the G395H data is shown in black, whereas the red line shows the best-fit combined model ($\rm BIC = 932$). Residuals to the model are shown in the middle panel.
    The fiducial fitting model is described in Sect.~\ref{sec:balmer_O3}.
    We indicate the masked regions based on the locations of possible narrow [\feii{}] emission in \citet{veron-cetty2004} (teal) and \ion{He}{i} $\lambda$6680 (purple). The [\ion{N}{ii}] component is not shown due to its relative flux being negligible. The bottom panel shows the \Halpha{} spectrum and best fit after subtracting the exponential component to highlight the P Cygni profile.}
    \label{fig:Ha_line_fit}
\end{figure}

\begin{figure}
    \centering
    \includegraphics[width=\linewidth]{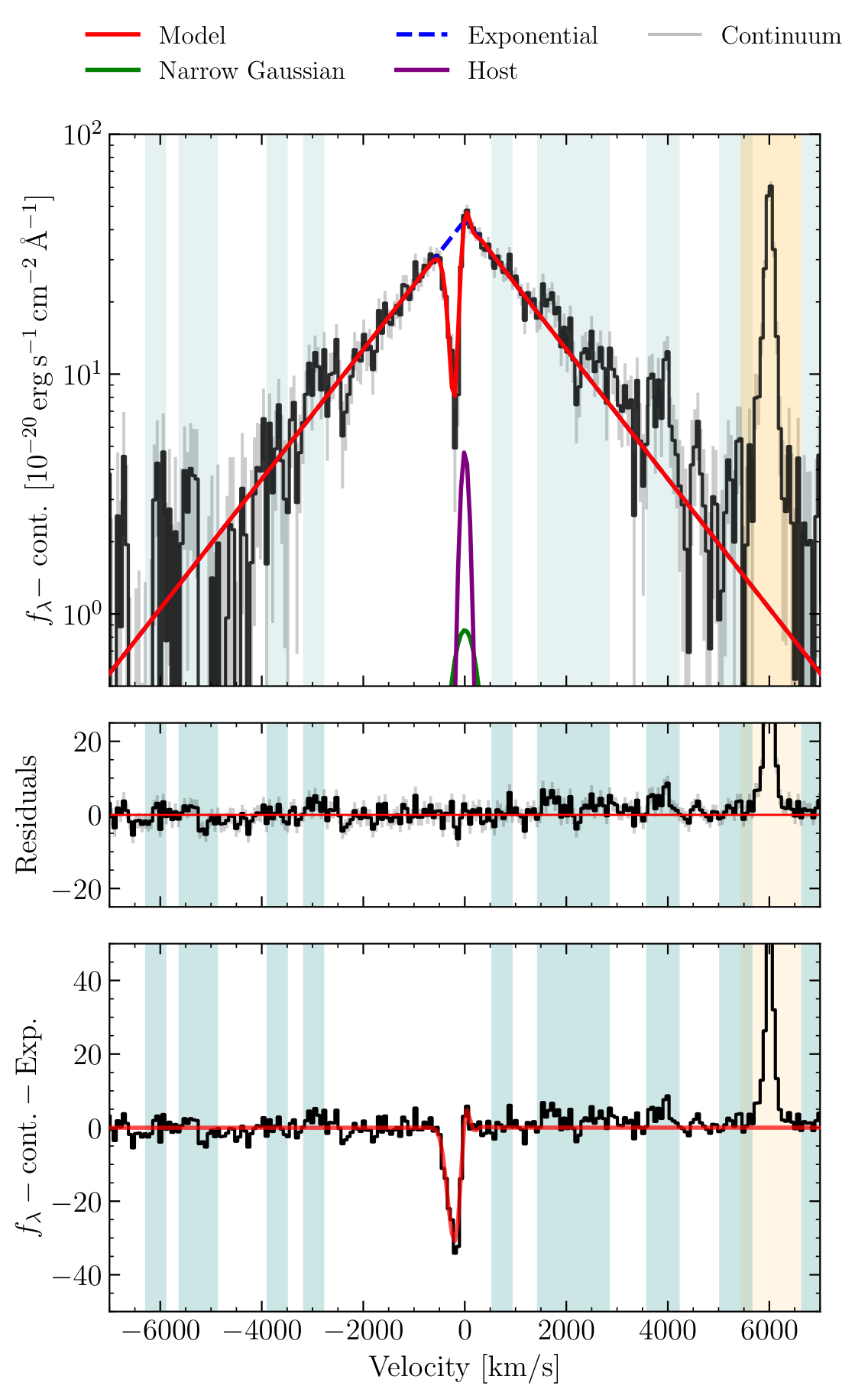}
    \caption{{{\bf H$\pmb\beta$  spectrum of GN-9771}. We used a similar model setup to that  for H$\alpha$ described in Fig.~\ref{fig:Ha_line_fit}} ($\rm BIC = 74$). The teal regions have been masked due to possible \feii{} emission. The orange region indicates the masked \oiii{} wavelengths. The bottom panel shows the \Hbeta{} spectrum and best fit (red) after subtracting the exponential component to highlight the P Cygni profile. Figure~\ref{fig:Hb_fit_fixed_exp} shows a version of the same fit with the exponential scale fixed to that fitted for \Halpha{}.}
    \label{fig:Hb_line_fit}
\end{figure}

To place our spectrum in the rest-frame, we obtain the redshift of the \oiii{} doublet of GN-9771 from the continuum-subtracted spectrum. This redshift will serve as reference for the rest of the analysis done in this paper because the redshift is most well defined due to the narrowness of the lines. The details of the continuum fitting and background subtraction of the grating spectrum are detailed in Appendix~\ref{sec:continuum_subtraction}. We fit two Gaussians with a fixed ratio of 2.98, and convolving the model with a Gaussian line spread function, using the nominal resolution of the G395H grating.\footnote{We obtain the Gaussian FWHM of the LSF as $\lambda/R$, with $\lambda$ being the observed wavelength of interest. The information about the NIRSpec dispersers resolution was taken from \url{https://jwst-docs.stsci.edu/jwst-near-infrared-spectrograph}.} We thus obtain a redshift of $z=5.53453 \pm 0.00004$, and a Gaussian width of FWHM~$202 \pm 5$~\unit{km.s^{-1}}. This narrow line width suggests a very low dynamical mass, similar to other LRDs with recent deep G395H observations \citep[e.g.,][]{deugenio2025, deugenio2025_absorber, wang2025}.

In Figs.~\ref{fig:Ha_line_fit} and~\ref{fig:Hb_line_fit} we show the high-quality H$\alpha$ and H$\beta$ spectra of GN-9771, respectively. The \Halpha{} spectrum confirms the absorption system seen in the NIRCam grism data \citep{matthee2024a} and we now show that the absorption is also present in H$\beta$ with a similar velociy offset. The total H$\beta$ flux is $\approx10.4$ times fainter than the H$\alpha$ flux which is a strong departure from the case B ratio of $\approx2.86$ (see Sect.~\ref{sec:balmer_lines_origin} for a discussion). The emission lines have prominent non-Gaussian wings \citep[see also][]{rusakov2025, degraaff2025}, whereas the core of the lines (the central $\pm 1000$~\unit{km.s^{-1}}) resemble P Cygni profiles.

We fit the \Halpha{} and \Hbeta{} lines of GN-9771 to a model consisting of a narrow Gaussian emission, a Gaussian absorber, a broad exponential, and a host Gaussian component. We call this our fiducial model.
These components are chosen arbitrarily to match the observed shape, and the individual physical interpretation of some of the components is not necessarily meaningful.
The host component has width and redshift fixed to those of the \oiii{} lines, is included in the model to represent a hypothetical host galaxy contribution.
The velocity offsets of all the components---except the host component---are allowed to vary within $\pm 500$~\unit{km.s^{-1}}.
For the fit, we mask the wavelengths potentially affected by [\feii{}]. We define the [\feii{}] masks using the lines seen in the N3 system of I Zw1 \citep{veron-cetty2004}. We also mask other detected lines such as \ion{He}{i} $\lambda$6680 and \oiii{} $\lambda\lambda 4960,5008$. Since the preliminary continuum subtraction based on a power law might not be completely accurate, we also introduce a wavelength-constant continuum level in the line fit.

We note that the [\feii{}] emission affects the flux of the exponential wings severely. The measured flux of [\feii{}] is $\approx 7$\% of the total \Hbeta{} flux within $\pm$3000~\unit{km.s^{-1}} of \Hbeta{}'s systemic velocity. This is only evident for this object due to the exceptionally high S/N of our spectrum. This component should be taken into account to investigate the differences in line profiles among different Balmer lines of other LRDs in future works.

\begin{table}
\centering
\caption{Table of best-fit parameters for \Halpha{} and \Hbeta.}\label{tab:Balmer_fitted_params}
\begin{tabular}{ccc}
\midrule
 & \Halpha{} & \Hbeta{} \\
\midrule
\multicolumn{3}{l}{{\bf Flux} (\unit{10^{-18}.erg.s^{-1}.cm^{-2}})}\\
\midrule
Narrow & 17.6 $\pm$ 0.9 & 0.5 $\pm$ 0.2 \\
Exponential & 199.1 $\pm$ 0.1 & 22 $\pm$ 2 \\
{[\ion{N}{ii}]} & 0.26 $\pm$ 0.12 & N/A \\
Host & 11 $\pm$ 12  & 6 $\pm$ 5 \\

\midrule
\multicolumn{3}{l}{\bf Absorption Optical depth ($\tau_0$)}\\
\midrule
Absorption & 1.24 $\pm$ 0.09 & 2.2 $\pm$ 0.6\\

\midrule
\multicolumn{3}{l}{\bf FWHM (\unit{km.s^{-1}})}\\
\midrule
Narrow & 330 $\pm$ 12 & 125 $\pm$ 34 \\
Absorption & 220 $\pm$ 5 & 303 $\pm$ 25 \\
Exp. & 1549 $\pm$ 5 & 2180 $\pm$ 490 \\
{[\ion{N}{ii}]} & 532 $\pm$ 92 & N/A \\
Host & 202 (fixed) & 202 (fixed) \\

\midrule
\multicolumn{3}{l}{\bf Vel. offset (\unit{km.s^{-1}})}\\
\midrule
Narrow & 136 $\pm$ 7 & -0.3 $\pm$ 13.0 \\
Absorption & -115 $\pm$ 3 & -155 $\pm$ 21 \\
Exponential & 41 $\pm$ 3 & 7 $\pm$ 31 \\

\bottomrule

\end{tabular}
\tablefoot{We recall that these values correspond to a physically motivated, yet purely empirical model. We list them here for reproducibility. The parameters of individual components might not have a direct physical interpretation. Exceptions to this are the widths of the exponential wings, which are much less degenerate with other components because  they extend over a broader range of wavelengths than the rest (see Figs.~\ref{fig:Ha_line_fit} and~\ref{fig:Hb_line_fit}).}
\end{table}

We list the parameters of the best-fitting models to \Halpha{} and \Hbeta{} in Table~\ref{tab:Balmer_fitted_params}.
The velocity offset of the exponential component implies that the exponential component is either not symmetric around the systemic redshift, or that there is a general velocity shift to the region that produces the \oiii{} emission.
The main narrow \Halpha{} emission component is significantly broader than the \oiii{} emission, while this is less conclusive for \Hbeta{}. We note that these components of our fiducial model are purely descriptive as they are meant to capture the general P Cygni shape of the line core. The specific physics involving resonant Balmer radiative transfer in Compton-thick media are complex and require more complex modeling \citep[e.g.,][]{shimoda2019, Chang25}.

The broad wings of the \Halpha{} line are very well described by a simple exponential up to at least $\sim\pm 8000$~\unit{km.s^{-1}}. Our fiducial model with exponential wings ($\rm BIC=932$) works significantly better than replacing the exponential with a broad Lorentzian component ($\rm BIC = 1027$, $\Delta\rm BIC \gg 10$; see Fig.~\ref{fig:Ha_line_fit_lorentz}). The best-fit exponential has a width of ${\rm FWHM_{exp, H\alpha}} = 1549 \pm 5$~\unit{km.s^{-1}}. For \Hbeta{} we obtain ${\rm FWHM_{exp, H\beta}} = 2180 \pm 490$~\unit{km.s^{-1}}. With this model, the best-fitting exponential wings of \Hbeta{} are wider than \Halpha{}, with a significance of $\sim$$1.3\sigma$. If electron scattering is the cause of the broad Balmer line wings \citep[e.g.,][]{rusakov2025}, the FWHM of the exponentials is expected to be the same as long as the scattering of the different lines is produced in the same layers within the envelope. This is due to the cross-section of the scattering being wavelength independent in the Thomson regime \citep[e.g.,][]{juodzbalis2025, brazzini2025,Chang25}. To test this, we both \Halpha{} and \Hbeta{} simultaneously, fixing the exponential width of both lines, and masking the central $\pm 1000$~\unit{km.s^{-1}}. Fixing the exponential to the same value yields an exponential width of $\rm FWHM = 1613 \pm 17$~\unit{km.s^{-1}}, and the best fit is not significantly worse ($\rm BIC = 258$) than leaving the widths of each line free ($\rm BIC = 251$).

To investigate the significance of the host Gaussian component, we repeat the fit to \Halpha{} and \Hbeta{} with our fiducial model, but removing this component. In the case of \Halpha{}, we obtain a best fit with $\rm BIC = 931$ ($\Delta\rm BIC\approx 1$; Fig.~\ref{fig:Ha_line_fit_no_host}), indicating that the host component has no statistical significance. In turn, for \Hbeta{} a better $\rm BIC = 67.6$ ($\Delta\rm BIC \approx -6$) is obtained. This result highlights the degeneracy of the model in the core of the line, as shown in Fig.~\ref{fig:Hb_line_fit_no_host}, our model with no host fits a narrow Gaussian component which could mimic a host \Halpha{} with a similar width to \oiii{}.

\begin{table}
\centering
\caption{List of relevant emission lines detected in the spectrum of GN-9771.}\label{tab:emlines}
\begin{tabular}{cccc}
\hline
Line & Flux$^a$ & FWHM$^b$ & $\Delta v^c$ \\
\hline
\Hgamma{} & $2.8 \pm 0.6$ & -- & -- \\
\oiii{} $\lambda 4364$ & $1.15 \pm 0.28$ & $361 \pm 74$ & $-15 \pm 5$ \\
\Hbeta{} & $19.8 \pm 0.6$ & $1866 \pm 106^e$ & -- \\
\oiii{} $\lambda 5008$ & $6.2 \pm 0.4$ & $202 \pm 5$ & 0\\
{[\feii{}]} opt. & $2.32 \pm 0.01$$^f$ & $464 \pm 61$ & $-43 \pm 15$ \\
\Halpha{} & $205.05 \pm 0.25$ & $1544 \pm 4^e$ & -- \\
\ion{He}{i} $\lambda 4472.74$ & $0.28 \pm 0.16$ & $320 \pm 100$ & $145 \pm 40$ \\
\ion{He}{i} $\lambda 4923.31$ & $0.21 \pm 0.09$ & $440$ (fixed) & $2.4$ (fixed) \\
\ion{He}{i} $\lambda 5017.08$ & $1.05 \pm 0.27$ & $440$ (fixed) &  $2.4$ (fixed)\\
\ion{He}{i} $\lambda 5877.25$ & $1.24 \pm 0.28$ & $440 \pm 85$ &  $2.4 \pm 1.4$ \\
\ion{He}{i} $\lambda 6680.00$ & $0.70 \pm 0.23$ & $395$ (fixed) &  $18$ (fixed) \\
\ion{He}{i} $\lambda 7067.14$ & $1.40 \pm 0.40$ & $395 \pm 90$ & $18 \pm 8$ \\
\hline
\end{tabular}
\tablefoot{
$^a$Line fluxes in~\unit{10^{-18}.erg.s^{-1}.cm^{-2}}.
$^b$FWHM of the emisison line in \unit{km.s^{-1}}.
$^c$Velocity with respect to \oiii{}.
$^d$Pseudo-continuum in the interval 2000--3600~\AA{}
$^e$Numerical FWHM is given, independently of any line fitting.
$^f$The flux is given for the band 5100-5400~\AA{}. Fluxes of individual [\feii{}] lines are given in Table~\ref{tab:feii_fluxes}.
}
\end{table}
\subsection{Balmer break} \label{sec:balmerbreak}

The rest-frame UV and optical spectrum of GN-9771 shows the characteristic features of many LRDs (see Fig.~\ref{fig:prism_spectrum}): a Balmer break, broad Balmer lines and a relatively blue UV continuum. The prism spectrum also shows forbidden lines with high excitation energies such as \oiii{} $\lambda\lambda$4960,5008, \oiii{} $\lambda$4364 (blended with H$\gamma$), [\ion{Ne}{iii}] $\lambda$3869 and \ion{C}{iii}] $\lambda$1909. Similar to A2744-45924, the prism spectrum shows indications of strong \feii{} emission, both in the UV and the optical (see below), and \ion{He}{i} and [\ion{O}{i}] emission. Notable as well are the absence of Ly$\alpha$ emission and the downturn in the far-UV $\lambda<2000$ \AA{} part of the spectrum, which is not seen in all LRDs.

The Balmer break is one of the key features in the spectra of LRDs \citep[e.g.,][]{setton2024}. We follow the parametrization of \cite{degraaff2025} and define it as $f_{\nu, 4000-4100}/f_{\nu, 3620-3720}$, where $f_{\nu, 3620-3720}$ is the median flux density over the $\lambda_0=3620$--$3720$~\AA{} range. For GN-9771, the Balmer break strength is $2.5 \pm 0.04$, which is close to the maximum value of $\approx3.0$ that typical stellar populations can reach \citep{wang2024} and similar to other very red LRDs (A2744-45294 has a break strength of $\approx3.4$). It is not as extreme as the Balmer breaks of ${\rm BB}\approx7$ measured in \textit{The Cliff} \citep{degraaff2025} or MoM-BH*-1 \citep{naidu2025}.

\subsection{A forest of {\rm[\feii{}]} emission, H$\gamma$, and {\rm\oiii{}} $\lambda 4634$}\label{sec:forb_feii_model}

\begin{figure*}
    \centering
    \includegraphics[width=\linewidth]{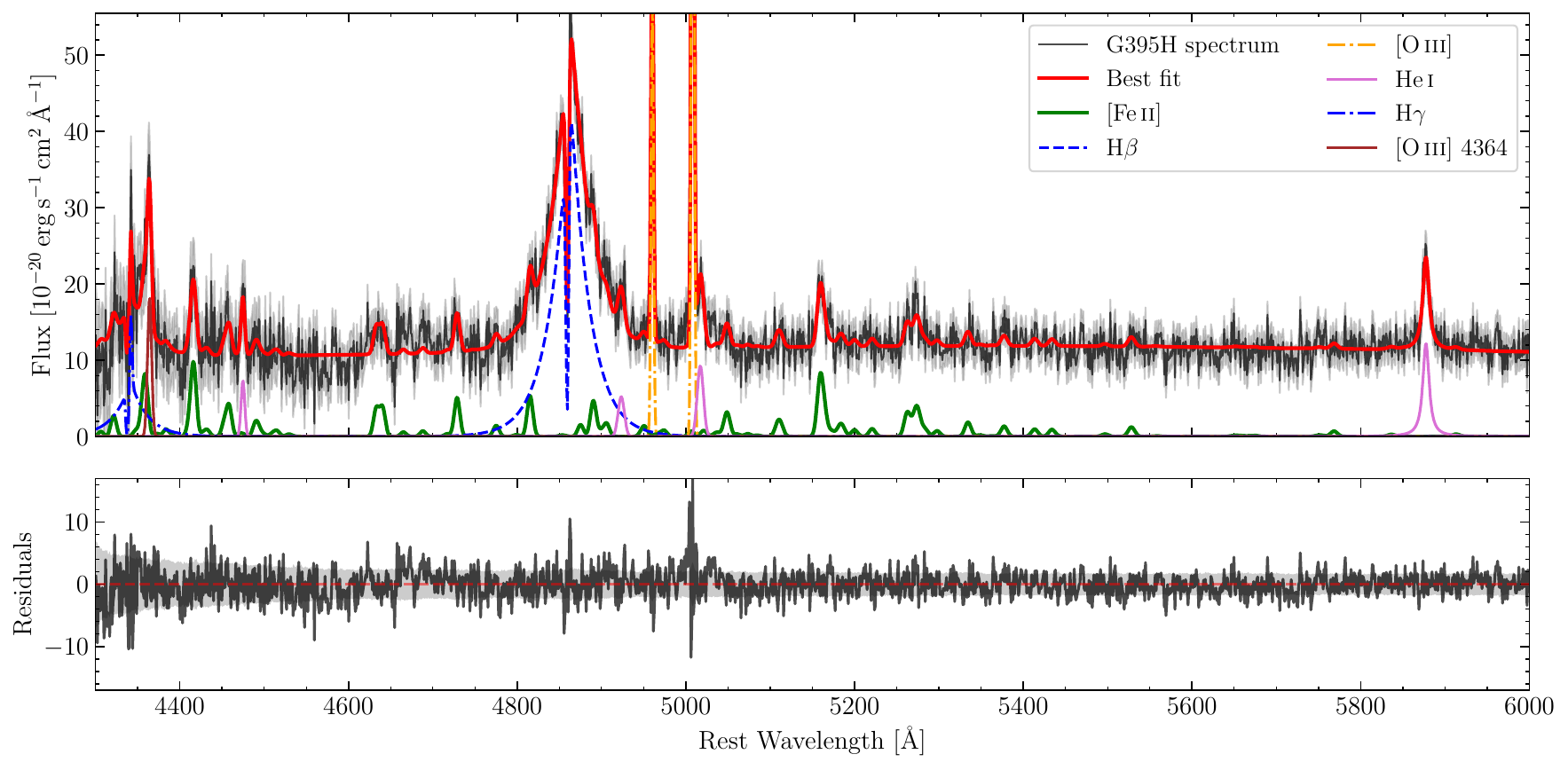}
    \caption{{\bf [\feii{}] forest detected in the G395H data over the $\bf\sim5000$ {\AA} region of GN-9771.} We show the fitted [\feii{}] spectrum as described in Sect.~\ref{sec:forb_feii_model} (green), the \ion{He}{i} lines (pink), \Hgamma{} (blue dot-dashed), the \Hbeta{} model (blue dashed), \oiii{} $\lambda\lambda 4960,5008$ (orange dot-dashed), and \oiii{} $\lambda4364$ (brown). The residuals of the fit as well as the observational uncertainties on the spectrum are shown in the bottom panel.}
    \label{fig:feii_fit_1}
\end{figure*}

Fig.~\ref{fig:feii_fit_1} shows our deep high-resolution grating spectrum zoomed in on the region around the \Hbeta{} and \oiii{} lines. Besides the very narrow \oiii{} emission discussed in Sect.~\ref{sec:balmer_O3} and the triangular \Hbeta{} profile, one can also notice a plethora of other fainter lines, most of which are emission lines from transitions from the singly ionized iron atom \feii{}. The lines around wavelengths 5200~\AA{} were initially identified as [\ion{Fe}{vii}] $\lambda$5276 and [\ion{Fe}{vii}] $\lambda$5159. However, our spectra do not show [\ion{Fe}{vii}] $\lambda$6087 which is usually stronger \citep{petrushevska2023, reefe2022}, nor lines from any other high-excitation iron transitions.
Upon inspecting the \feii{} templates from I Zw1, we realized that most of the features were in fact unusually narrow [\feii{}] emission lines.
 
Some of the most prominent [\feii{}] emission lines lie very close to the rest-frame wavelength range of the broad \Hbeta{} and the \oiii{} doublet. In particular, the line profile of the broad wing of \Hbeta{} is contaminated by the relatively strong [\feii{}] emission-line at 4815.9~\AA{}. Before fitting the [\feii{}] complex, we subtract the continuum as well as the model of \Hbeta{} and \Halpha{}, which were fit after masking all the potential [\feii{}] lines as described in Sect.~\ref{sec:balmer_O3}. The full model is convolved with the wavelength-dependent line-spread function, as defined by the nominal (wavelength-dependent) resolution of the G395H grating.

We obtain the list of forbidden (magnetic-dipole and electric-quadrupole) \feii{} transitions in the range 4200--8000~\AA{} from the NIST\footnote{\url{https://www.nist.gov/pml/atomic-spectra-database}} database \citep{akramida2024}.
Within a forbidden transition multiplet, the relative strength of the component $n$ can be obtained as
\begin{equation}
    r_n \propto (2 J_{i} + 1)\,\lambda^{-1}\,A_{ki} \,, \label{eq:feii_forbidden_ratios}
\end{equation}
where $J_i$ is the quantum number of the total angular momentum relative to the lower level, $\lambda$ the emitted vacuum wavelength, and $A_{ki}$ the Einstein coefficient of the transition.

We fit our [\feii{}] model to the rest-frame wavelength ranges 4200--6000~\AA{} and 6350--8000~\AA{}, which are the two wavelength intervals of our data separated by the detector gap. These first interval contains \Hgamma{} and \Hbeta{}, and the second interval includes \Halpha{}. The fitting model consists of a power-law continuum, the \Hbeta{} or \Halpha{} models as described above, a double Gaussian for the \oiii{} doublet with fixed flux ratio of 2.98, and the [\feii{}] model. The former consists of Gaussians at the positions of every forbidden line in the NIST database, with a shared velocity width. We also allow for a small velocity offset as a free parameter, shared by all the [\feii{}] lines. The Gaussian amplitudes within every multiplet are fixed using Eq.~\ref{eq:feii_forbidden_ratios}. In addition, we include in our model the \ion{He}{i} $\lambda$4472, $\lambda$4923, $\lambda$5017, and $\lambda$6680 as single Gaussians whose amplitudes and widths can vary freely, and the brighter \ion{He}{i} $\lambda$5876 and $\lambda$7067 as Lorentzians. In the case of these two lines, a Lorentzian gives a significantly better fit than a Gaussian, hinting a broad component. The \ion{He}{i} spin-singlet lines at $\lambda$4923 and $\lambda$5017~\AA{} are blended with \Hbeta{} and \oiii{}, respectively. For this reason, we fixed the widths and velocity shifts of these two lines to the values fit for \ion{He}{i} $\lambda$5876.\footnote{The Gaussian FWHM is matched with the Lorentzian FWHM.}
We also include \Hgamma{} and \oiii{} $\lambda4364$ components in the fit. The wavelengths of these two lines are very close to each other, which difficultates a precise assesment of both, moreover, strong [\feii{}] emission is also expected in nearby wavelengths. For this reason, we model for \Hgamma{} consisting in a broad exponential with fixed velocity width to that fit to \Hbeta{}, a Gaussian absorber and a Gaussian host component with width tied to \oiii{} $\lambda\lambda 4960,5008$.

\begin{figure}
    \centering
    \includegraphics[width=\linewidth]{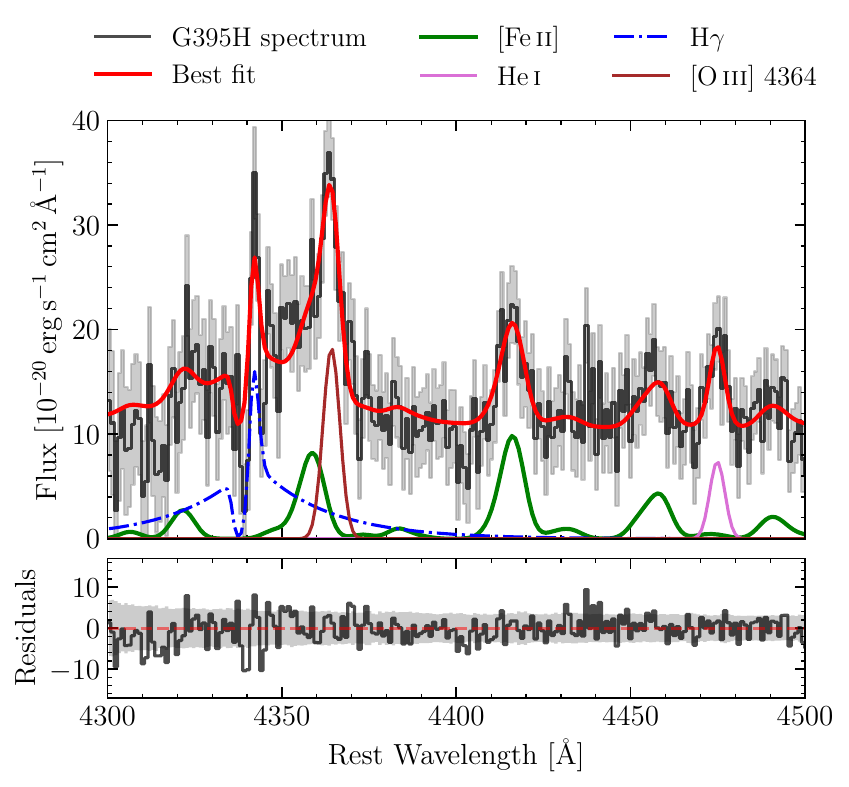}
    \caption{Same as Fig.~\ref{fig:feii_fit_1}, but zoomed in on the wavelength range containing H$\gamma$ and \oiii{} $\lambda 4364$.}
    \label{fig:feii_fit_1_zoom}
\end{figure}

\begin{table}
\centering
\caption{Main fitted forbidden [\feii{}] line fluxes of multiplets.}\label{tab:feii_fluxes}
\begin{tabular}{ccc}
\hline
Transition & Main line [\AA{}] & Flux$^\dag$\\
\hline
a$^{4}$G-a$^{4}$F & 4320.84 & 3.20 $\pm$ 1.05 \\
b$^{2}$F-a$^{4}$D & 4323.14 & 1.03 $\pm$ 0.38 \\
b$^{2}$P-a$^{4}$F & 4357.37 & 11.67 $\pm$ 3.26 \\
a$^{6}$S-a$^{6}$D & 4360.56 & 18.51 $\pm$ 6.99 \\
b$^{4}$F-a$^{6}$D & 4417.51 & 60.30 $\pm$ 13.51 \\
b$^{2}$D-a$^{4}$P & 4453.71 & 5.71 $\pm$ 2.07 \\
a$^{4}$H-a$^{6}$D & 4633.57 & 32.47 $\pm$ 8.51 \\
b$^{4}$P-a$^{6}$D & 4729.39 & 43.27 $\pm$ 10.58 \\
b$^{4}$F-a$^{4}$F & 4815.89 & 47.11 $\pm$ 12.45 \\
b$^{2}$D-a$^{2}$G & 4899.97 & 5.24 $\pm$ 1.78 \\
a$^{2}$D$^{2}$-a$^{6}$D & 4967.18 & 3.19 $\pm$ 1.44 \\
c$^{2}$D-a$^{2}$P & 5049.59 & 27.43 $\pm$ 7.82 \\
a$^{4}$H-a$^{4}$F & 5160.23 & 48.43 $\pm$ 10.37 \\
a$^{2}$F-a$^{4}$D & 5165.40 & 20.59 $\pm$ 7.60 \\
b$^{4}$P-a$^{4}$F & 5274.83 & 30.50 $\pm$ 7.29 \\
a$^{2}$H-a$^{4}$F & 5281.74 & 0.77 $\pm$ 0.38 \\
a$^{2}$D$^{2}$-a$^{4}$F & 5528.89 & 13.17 $\pm$ 3.97 \\
c$^{2}$D-b$^{4}$P & 5769.12 & 4.63 $\pm$ 1.83 \\
c$^{2}$G-a$^{2}$G & 5837.07 & 3.03 $\pm$ 0.98 \\
b$^{2}$G-a$^{4}$P & 5912.23 & 1.39 $\pm$ 0.52 \\
c$^{2}$D-a$^{2}$D$^{2}$ & 5914.88 & 1.75 $\pm$ 0.72 \\
\hline
\end{tabular}
\tablefoot{
$^\dag$Flux units are~\unit{10^{-20}.erg.s^{-1}.cm^{-2}}.
}
\end{table}

We find that the [\feii{}] lines are at a similar velocity as the \oiii{} $\lambda\lambda 4960, 5008$ velocity (offset by $-43 \pm 15$~\unit{km.s^{-1}}). The [\feii{}] lines have an intermediate width of ${\rm FWHM} = 464 \pm 61$~\unit{km.s^{-1}}, comparable to the width of the \ion{He}{i} lines (within 1$\sigma$; see Table~\ref{tab:emlines}), significantly broader than the \oiii{} $\lambda\lambda$4960,5008 doublet with ${\rm FWHM} = 156 \pm 2$~\unit{km.s^{-1}}, but much narrower than the Balmer lines. 
The best-fitting model is shown in Figs.~\ref{fig:feii_fit_1} and~\ref{fig:feii_fit_1_zoom}, and individual fluxes of the main ($\rm S/N > 1$) [\feii{}] lines are listed in Table~\ref{tab:feii_fluxes}.

The width of the best-fitting \oiii{} $\lambda 4364$ Gaussian is $361\pm 74$~\unit{km.s^{-1}}, seemingly broader than the \oiii{} $\lambda\lambda 4960,5008$ conterparts (with $\rm FWHM = 202 \pm 5$~\unit{km.s^{-1}}), but with a significance of only 2$\sigma$. As shown in Fig.~\ref{fig:feii_fit_1_zoom}, \oiii{} $\lambda 4364$ is potentially affected by a strong [\feii{}] line, hindering a precise assesment. In fact, fixing the width of this Gaussian component to that of the $\lambda\lambda 4960,5008$ doublet does not significantly worsen the fit ($\rm\Delta BIC \approx 4$). We discuss the implications of the \oiii{} line width and the \Hgamma{} fluxes in Sect.~\ref{sec:host_implications}.

\subsection{UV {\rm\feii{}}: Absorption or emission}\label{sec:uv_feii}

The emission spectrum of \feii{} is complex, having hundreds of significant transitions and spanning from the UV to the optical and near infrared parts of the spectrum. We investigate the UV \feii{} emission of GN-9771. \citet{labbe2024} first reported prominent UV \feii{} emission in their deep spectrum of the luminous LRD A2744-45924. However, the low resolution ($R\sim 100$) of the PRISM spectrum hinders a assessment of the multiple \feii{} transitions. GN-9771 presents a very similar UV \feii{} bump, as shown in Fig.~\ref{fig:prism_spectrum}.

We fit the UV \feii{} in the range 2000-3600~\AA{}, using templates based on the galaxy I Zw1 \citep{vestergaard2001, salviander2007}. In particular, we use the version of these templates from the code PyQSOFit \citep{guo2018}, and convolving the model with an arbitrary Gaussian, to include the effects of the PRISM resolution and the intrinsic velocity dispersion of the lines. Due to the huge number of transitions, we do not attempt to fit a deconvolved width for UV \feii{}. Following \citet{labbe2024}, we also include the \ion{Mg}{ii} $\lambda$2799 and \ion{He}{ii} $\lambda$3203 as Gaussians in the model.

\begin{figure}
    \centering
    \includegraphics[width=\linewidth]{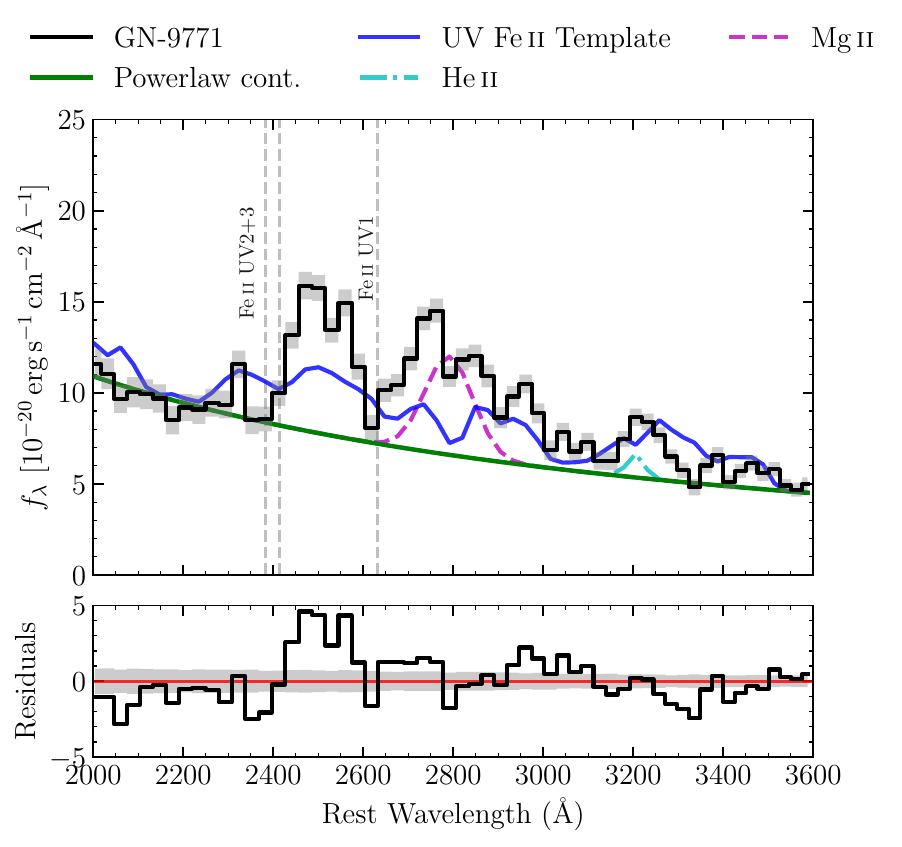}
    \caption{{\bf Template fit to UV \feii{}.} We fit templates to the UV pseudo-continuum emission. The fitting is challenging due to the limited PRISM resolution in this wavelength range ($R\sim50$). The model includes the \ion{Mg}{ii} and \ion{He}{ii} lines, which are usually strong. The template fitting does not produce a satisfactory result ($\chi^2_\nu = 7.5$), in particular for the $\sim$2500~\AA{} bump. Absorption features from the resonant \feii{} multiplets UV1 and UV2+UV3 are compatible with the observed shape of the continuum, which are also predicted by our Cloudy models (see Sect.~\ref{sec:host_implications} for a discussion).}
    \label{fig:UV_feii_fit}
\end{figure}

The template fitting does not provide a satisfactory result ($\chi^2_\nu=7.5$; Fig.~\ref{fig:UV_feii_fit}), particularly failing to reproduce the prominent bump at $\sim$2500~\AA{}. The secondary bump at $\sim$2800~\AA{} can only be matched by invoking a strong contribution from broad \ion{Mg}{ii} \citep[see also][]{labbe2024, tripodi_deep_2025}. An alternative explanation would be that the $\sim$2500~\AA{} bump could arise from absorption by the resonant \feii{} multiplets UV1 and UV2+UV3, at $\sim$2600 and $\sim$2400~\AA{}, respectively (see Sect.~\ref{sec:host_implications} for a detailed discussion).

\subsection{{\rm\ion{He}{i}} emission}\label{sec:HeI}

As described in Sect.~\ref{sec:forb_feii_model}, we also identify six \ion{He}{i} lines in our spectrum. Despite the high resolution of the G395H grating, some \ion{He}{i} lines are blended with the broad tails of the \Hbeta{} and \Halpha{} lines. We first fit the \ion{He}{i} line $\lambda$5876---which has the highest S/N---, with a Lorentzian profile, which gives a slightly better $\chi^2_\nu$ than a Gaussian. Then, we fit the \ion{He}{i} $\lambda$5017 and $\lambda$4923 with Gaussians of the same FWHM as the Lorentzian \ion{He}{i} $\lambda$5876, and matching their velocity offset. The \ion{He}{i} $\lambda$4472 line is fitted independently with a Gaussian. The other high S/N line \ion{He}{i} $\lambda$7067 is also fitted to a Lorentzian, and the blended \ion{He}{i} $\lambda$6680 fit as a Gaussian, fixing the width and velocity offset in the same manner. Interestingly, \ion{He}{i} $\lambda$7067 is the strongest helium line that we cover, which is typically much weaker than \ion{He}{i} $\lambda$5877 in normal galaxies, and it is indicative of high gas density (see Sect.~\ref{sec:hei_origin} for a discussion).

In Table~\ref{tab:emlines} the fluxes of all the \ion{He}{i} lines and others are listed. Both \ion{He}{i} $\lambda$5876 and $\lambda$7067 do not show a significant velocity shift with respect to \oiii{} $\lambda\lambda 4960,5008$ ($\Delta v = 2.4 \pm 1.4$ and $18 \pm 8$~\unit{km.s^{-1}}, respectively). \ion{He}{i} $\lambda$4472 appears to be redshifted with respect to \oiii{} by $\Delta v = 145 \pm 40$~\unit{km.s^{-1}}, but the low $\rm S/N\approx 2$ of this line makes this measurement unreliable.
We also detect remarkably strong \ion{He}{i} $\lambda$4923, $\lambda$5017 and $\lambda$6680 singlet lines. We note that there are [\feii{}] multiplets contaminating these lines (especially the first two); however, our model disfavors them when the theoretical [\feii{}] line ratios are taken into account.

\section{Photoionization modeling} \label{sec:phot_models}

In this section we present Cloudy modeling to interpret the observed spectral features. We caution that these models are empirically motivated phenomenogical models whose main goal is to provide qualitative insights into the conditions and structure of the gas around LRDs.

\subsection{Photoionized cloud model}\label{sec:cloudy_setup}

We use Cloudy \citep[version 23.01, last described in][]{chatzikos2023, gunasekera2023} to simulate an intrinsic ionizing SED striking a cloud of very dense gas. The incident SED is that of a classical AGN with a blackbody temperature of $10^5$~K, and power-law slopes $\alpha_{\rm OX} = -1.5$, $\alpha_{\rm UV}=-0.5$, and $\alpha_{\rm X}= -0.5$ \citep[][]{jin2012}. We use the default atomic data for all the species, and solar relative abundances. These parameters are comparable to those used in similar setups to investigate the Balmer break of LRDs \citep{naidu2025, degraaff2025, ji2025}. Similar models were explored thoroughly in the literature to investigate the \feii{} emission of AGN \citep[e.g.,][]{baldwin1995, baldwin2004, wang2008, ferland2009, sameshima2011, sarkar2021}. We set the normalization of the incident SED varying the ionization parameter $\logten U$ from $-4$ to 0 with 1 dex steps. We simulate clouds with plane-parallel geometry, in a fairly broad range of hydrogen density $\Hden = 10^6$ to $10^{14}$~\unit{cm^{-3}} and hydrogen column density $\NH=10^{21}$ to $10^{26}$~\unit{cm^{-2}}, both with 1 dex steps. We also vary the turbulence velocity $v_{\rm turb}=10$ to 510~\unit{km.s^{-1}} in steps of 100~\unit{km.s^{-1}} and the metallicity $\logten (Z/Z_\odot)=-2$ to 1 with 1 dex steps. This produces a grid of 6480 models. We note that variations in the incident spectrum are to some extent degenerate with variations in the properties of the dense gas and therefore merit exploration on their own \citep[e.g.,][]{BWang25}. However, we have verified that our qualitative results on the density-temperature structure and origin of observed emission lines are not strongly impacted by our specific choice. As such variations are also degenerate with the possible contribution from the host galaxy, we defer the detailed investigation where the incident spectrum is also varied to the future when better  high-resolution data is available in the UV and blue optical regime.

We fit the observed rest-frame optical spectrum with the \feii{} emission predicted by the Cloudy models. The templates, convolved with a Gaussian---free width, velocity offset, and normalization---are combined with the same continuum and line components as in Sect.~\ref{sec:forb_feii_model}. The fit is performed on the relative \feii{} fluxes rather than equivalent widths, focusing on the line flux ratios.

\begin{figure}
    \centering
    \includegraphics[width=\linewidth]{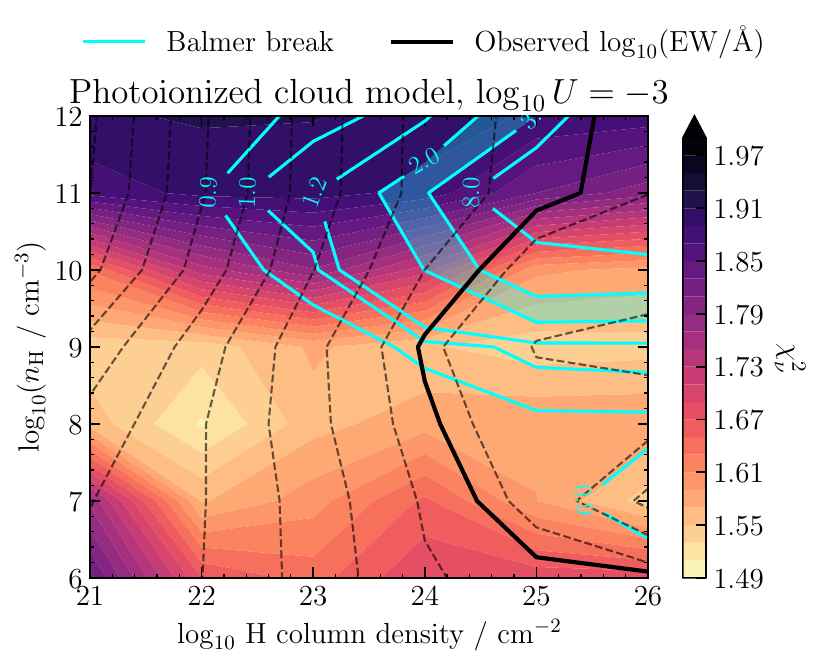}
    \caption{{\bf Density and column density of the warm layer as preferred by the [\feii{}] line-emission compared to that preferred by the Balmer break strength.} The colored contours show the values of $\chi^2_\nu$ of our fit to the optical [\feii{}] of GN-9771 to Cloudy photoionized cloud models with $\vturb = 210$~\unit{km.s^{-1}} and $Z/Z_\odot = 0.01$. We also show the strength of the Balmer break for the grid of  models as cyan contour lines, where we highlight the approximate Balmer break strength ($\approx$2.5) of GN-9771 with a shaded region. The black line marks the contour of the models with matching [\feii{}] EW to the observed in GN-9771, in the range 5100--5400~\AA{}. Dashed lines indicate contours of $\times 2$ difference in EW (increasing toward the right-hand side of the figure).}
    \label{fig:BHS_chisquare_contours}
\end{figure}

We perform a least-squares fit of each template of the grid to the rest-frame optical \feii{} spectrum of GN-9771---no other measurements are included in this fit. We find that the best fit ($\chi^2_\nu = 1.49$) is obtained for the model with $\logU = -3$, $\Hden = 10^8$~\unit{cm^{-3}}, $\NH = 10^{22}$~\unit{cm^{-2}}, $\vturb = 210$~\unit{km.s^{-1}} and $\logten{Z/Z_\odot}=-2$.
In Fig.~\ref{fig:BHS_chisquare_contours} we show the minimum $\chi^2_\nu$ across the grid for each point in the (\Hden{}, \NH{}) plane for this photoionized cloud model. In general, this model reproduces relatively well the observed optical [\feii{}] for densities of $\Hden = 10^\text{7--10}$, and slightly prefers a relatively low column density.
Lower values of ionization parameter (i.e., $\logU = -4$) yield similarly good fits to the [\feii{}] line ratios of GN-9771, but have more difficulties to reproduce the strong Balmer break at the same time. The metallicity parameter, in turn, has a minor role for line ratios, yielding similar results for all probed values ($\log_{10}(Z/Z_\odot) = -2$ to 0). The turbulence parameter $\vturb$ also has a minor effect on the optical [\feii{}] ratios; however, a high \vturb{} is preferred to obtain a smooth Balmer break \citep[see also][]{ji2025,naidu2025}.

In Fig.~\ref{fig:BHS_chisquare_contours} we also show the Balmer break strength produced by the models in the grid. Only the models with $\Hden \approx 10^\text{9--10}$~\unit{cm^{-3}} and $\NH \gtrsim 10^{24}$~\unit{cm^{-2}} can produce reasonable [\feii{}] line ratios---taking into account the simplicity of the model---and simultaneously accomodate Balmer breaks that roughly match the one observed in GN-9771. In any case, this model is only able to reproduce the [\feii{}] for $\Hden < 10^{10}$~\unit{cm^{-3}}.
We conclude that the photoionized cloud model prefers gas densities of $\Hden =10^\text{9--10}$~\unit{cm^{-3}} and high column densities $\NH>10^{24}$~\unit{cm^{-2}} to reproduce the observed [\feii{}] ratios and Balmer break of GN-9771, as well as the EW of the [\feii{}] lines.
On the one hand, the EW of the [\feii{}] emission increases for higher values of \NH{}, highlighting the need of a thick layer of gas adding to the total [\feii{}] emissivity. On the other hand, the EW also naturally increases for higher gas metallicity, hence there is a degeneracy between the two parameters.
One caveat of the choice of a plane-parallel geometry is that this assumption is only valid in the case that the distance from the ionizing source to the inner face of the cloud is much larger than the cloud thickness. For our models, the inner radius is set by the ionization parameter, if one would assume a spherical geometry. For $\logU = -3$ and $\logten(\Hden /\rm cm^{-2}) = 9$--10 the inner radius would be $\approx 10^{17}$--$10^{18}$~\unit{cm}, roughly more than two orders of magnitude larger than the implied cloud depth given by $\NH/\Hden$.

\subsection{Overall spectral shape}\label{sec:balmer_break_origin}

In photoionized cloud models (BH*), the Balmer break strength is explained as the absorption of the intrinsic (AGN-like) SED by the partially ionized gas. The strength of the break is thus directly suggestive of high gas column densities \citep[e.g.,][]{inayoshi2025, naidu2025}. The Balmer break strength in this scenario also depends on the gas volumetric density and temperature \citep[e.g.,][]{liu2025} and the dense gas covering fraction. 

\begin{figure}
    \centering
    \includegraphics[width=\linewidth]{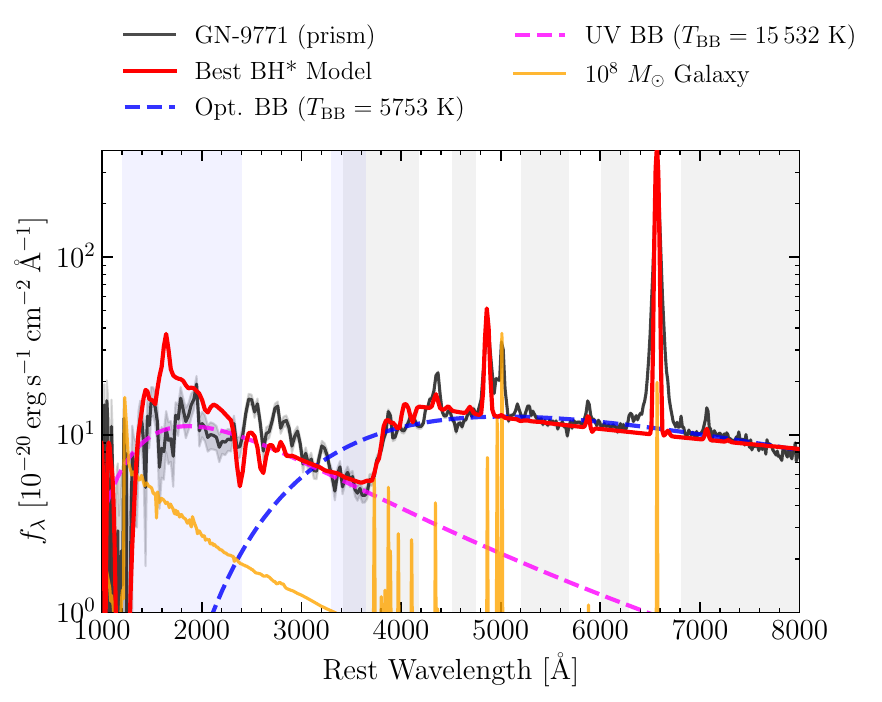}
    \caption{{\bf Best-matching BH* model and blackbody fit to the optical spectrum.} We matched the gray shaded wavelengths to the larger model grid presented in \citet{naidu2025}. We also show the blackbody that  best fits  the rest-frame optical (dashed blue) and UV (dashed purple), for which we obtain an effective temperature of $T_{\rm BB, opt}=5753$~K (fit to wavelengths highlighted with a gray shade) and $T_{\rm BB, opt}=15\,530$~K (purple shade), respectively. For illustration, we also include the SED fit to the object ALT-31334 ($z=5.66$; $M_* = 10^8\ M_\odot$), a typical compact blue galaxy that could resemble a hypothetical host for GN-9771 (see Sect.~\ref{sec:host_implications}).}
    \label{fig:prism_w_cloudy_BHS}
\end{figure}

In Fig.~\ref{fig:prism_w_cloudy_BHS}, we show the best-fitting BH* continuum to the PRISM spectrum of GN-9771, from a larger grid of models presented in \citet{naidu2025}, extended down to $\logU = -4$.
This larger sample also explores variations in the input SED parameters, so we adopt it for fitting the continuum.
A pre-selection was made to match the approximate \Halpha{} and \Hbeta{} EWs ($\rm EW(\Halpha) = 1500$--2000~\AA{}, $\rm EW(\Hbeta)=100$--200~\AA{}) and Balmer break (0.5--5).
The best-fitting model is selected by least-squares minimization matching the continuum to the GN-9771 prism data in the shaded gray wavelength intervals shown in Fig.~\ref{fig:prism_w_cloudy_BHS}, adding an arbitrary normalization.
The best-fit model corresponds to an incident AGN SED with $T_{\rm BB}=5\times10^4$~K, $\alpha_{\rm OX}=-1$, $\alpha_{\rm UV} = -1$, $\alpha_{\rm X}=-0.5$, $\logU = -3$, $\Hden = 10^{10.5}$~\unit{cm^{-3}}, $\NH = 10^{23.5}$~\unit{cm^{-2}}, $\vturb = 300$~\unit{km.s^{-1}} and $Z/Z_\odot=0.01$.
The best-matched BH* model is convolved with the PRISM line-spread function. The BH* model is capable to reproduce the overall shape of the optical+UV continuum of GN-9771, including the shape of the Balmer break due to turbulent broadening of the Balmer series absorptions.
The fact that the best-fit parameters are similar to those obtained in Sect.~\ref{sec:cloudy_setup} suggests that a BH* model can simultaneously reproduce both the overall spectral shape and optical [\feii{}] emission, although a more definitive assessment is left for future work.
The lack of some emission lines, for example \oiii{} and [\ion{Ne}{iii}] $\lambda$3869, could be explained by an additional host galaxy component, due to their low critical densities (further discussed in Sect.~\ref{sec:host_implications}).

For reference, we also fit a Planck blackbody law to the optical and UV continua of GN-9771 \citep[see, e.g.,][]{zwick2025}. The optical (3640--8000~\AA{}) continuum is well fit by a blackbody with an effective temperature of $T_{\rm BB, opt}=5753$~K, using the wavelength ranges marked as shaded gray areas in Fig.~\ref{fig:prism_w_cloudy_BHS}. Despite the good fit over a broad range of optical wavelengths, the spectrum of GN-9771 has a more abrupt drop-off than a blackbody toward the Balmer break ($\approx$3700~\AA{}), this suggests that a blackbody alone cannot reproduce the emission, but a Balmer break from neutral hydrogen absorption is also needed in this case. The measured blackbody temperature of the optical emission is in line with \citet{liu2025}, who suggested that the continuum emission of LRDs could be explained as a cool photosphere with $T\approx 5000$~K. Our BH* models can reach similarly low temperatures at the edge of their envelopes (see Fig.~\ref{fig:BHS_Te_ne}). Some works have proposed that the UV emission of LRDs could arise from hotter accretion disk components \citep[e.g.,][]{inayoshi2025_binaries, zhang2025}. We also fit another blackbody to the UV continuum (1000--3600~\AA{}), with an effective temperature of $T=15\,530$~K. We note that our BH* models can mimic such blackbody-like continuum shape in the UV as the result of the absorption of the incident UV powerlaw by the neutral gas.
However, we note that adding these two blackbodies would not reproduce the sharpness of the Balmer break, yielding a profile that is too smooth \citep[e.g.,][]{setton2024, wang2024, ma2025}.

The BH* model can reproduce the very high \Halpha{}/\Hbeta{} ratio, and also shows very weak H$\gamma$ emission.
The \oiii{} lines are not well reproduced by our BH* model ($\lambda$4364 and $\lambda\lambda$4960,5008), but these could be explained with contributions from a host galaxy (see Sect.~\ref{sec:host_implications}).

\section{Interpretation} \label{sec:interpretation}

An attractive feature of photoionized slab models \citep[from now on, also referred to as \textit{Black Hole Star} models; BH*;][]{naidu2025} is its capability to explain many of the most unusual spectral features of LRDs with the simple approach of a constant density gas illuminated by an incident ionizing SED. Therefore, we use this phenomenological model to interpret the origin of various observed spectral features in the spectrum of GN-9771. Table~\ref{tab:emission_features} summarizes the key points and challenges.

\subsection{Where the various spectral features originate}\label{sec:emline_origin}

\begin{figure}
    \centering
    \includegraphics[width=\linewidth]{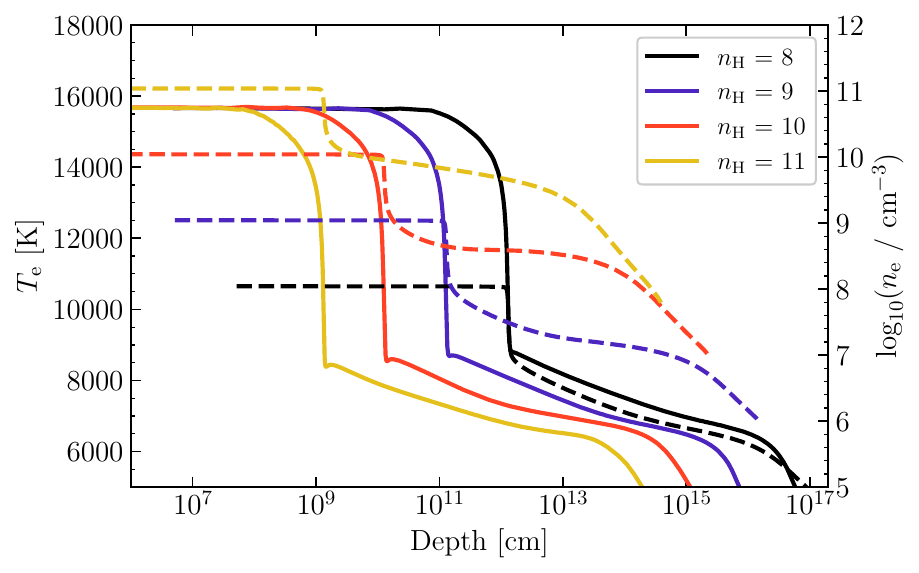}
    \caption{{\bf Temperature-density structure of a suite of BH* models.}  We show the electron temperature (solid lines) and electron density (dashed lines) profiles for models corresponding to $\logten U = -3$, $\vturb = 210$~\unit{km.s^{-1}} and $Z / Z_\odot = 0.01$, for a wide range of hydrogen densities. All models are characterized by a hot ($\sim16\,000$~K; see Fig.~\ref{fig:BHS_Te_logU} for the temperature with different choices of  ionization parameter) and dense ionized region whose temperature and density abruptly drop at an ionization front, its radius dependent on density, and a thermalized outer layer of $\sim6000$--$7500$~K.}
    \label{fig:BHS_Te_ne}
\end{figure}

In photoionized cloud (BH*) models, due to the high densities considered, most UV ionizing radiation is deposited in a thin hot layer at the illuminated face of the cloud, generating an ionization front.
The gas in the outer layers is then thermalized to temperatures of 6000--7500~K.
In Fig.~\ref{fig:BHS_Te_ne} we show the temperature and electron density dependence for BH* models with $\logU = -3$ and varying density (see also Fig.~\ref{fig:BHS_Te_logU} for the dependence of these on the choice of \logU{}).
The thermalized layer is partially ionized, that is, there is still a significant amount of free electrons ($\eden / \Hden \sim 0.01$--0.001). The size of this region depends mainly on the density and the ionization parameter, while its temperature is stable with the choice of different ionization parameters for the same density. The metallicity also plays a role in the total size of the thermalized region, as it regulates the temperature drop in the outer layers, and \vturb{} has a negligible effect.

\begin{figure}
    \centering
    \includegraphics[width=\linewidth]{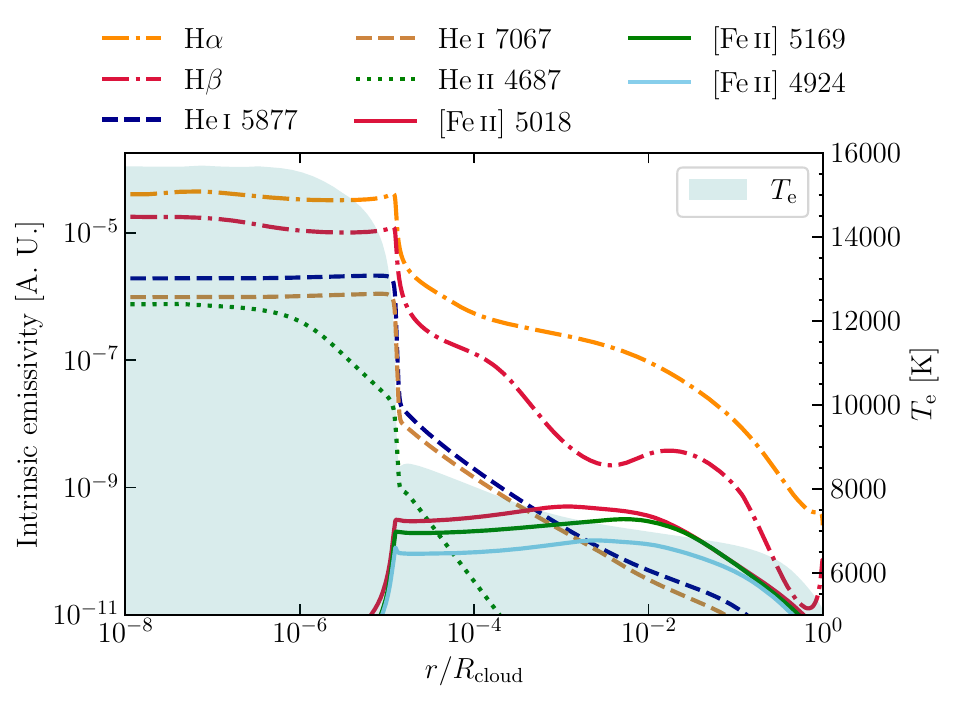}
    \caption{{\bf Origins of the lines.} The intrinsic emissivity of selected emission lines in the photoionized Cloudy model, for $\Hden = 10^{10}$~\unit{cm^{-3}}, and $\NH = 10^{25}$~\unit{cm^{-2}}. Emissivities are expressed in arbitrary units of energy per unit time per unit volume. The ionization parameter is $\logU = -3$. The hydrogen Balmer and helium lines are intrinsically strong in the inner layers due to recombination emission, although the opacity of \Halpha{} and \Hbeta{} is high, which leads to scattering effects. [\feii{}] is mostly emitted in a warm layer with $\Te=6000$--7500~K, which is shielded from the far-UV ionizing radiation.}
    \label{fig:BHS_line_emissivities}
\end{figure}

In Fig.~\ref{fig:BHS_line_emissivities} we show the outward emissivity of some optical emission lines for BH*-like models with $\Hden = 10^{10}$~\unit{cm^{-3}}, for a total column density of $\NH = 10^{25}$~\unit{cm^{-2}}. 
The photon energy needed to ionize an iron atom to the Fe$^+$ is 7.9~eV, and those are ionized further with photon energies exceeding 16.5~eV. For comparison, the ionization potential of neutral Hydrogen is 13.6~eV. 
The forbidden [\feii{}] lines have relatively low critical densities of \citep[$\eden \approx 10^\text{6--7}$~\unit{cm^{-3}} at $\Te = 10^4$~K, weakly dependent on temperature;][]{mendoza2023}.
[\feii{}] is thus characteristic of partially ionized regions \citep[e.g.,][]{zhang2024}. 
However, we note that in gas with high column density, significant line emission can still arise at electron densities well above the critical value, since the contributions from a large path length of emitting material along the line of sight add up.
In our photoionized models, the optical [\feii{}] emissivity is maximal for a gas temperature of $\sim 7500$~K.

We conclude that the strong [\feii{}] emission lines of GN-9771 arise from a warm ($\Te\approx 7000$~K) outer layer of dense ($\Hden \approx 10^{9\text{--}10}$\unit{cm^{-3}}) and partially ionized ($\eden \approx 10^{7\text{--}8.5}$\unit{cm^{-3}}) gas with high column density ($\NH\approx 10^{24}$~\unit{cm^{-2}}). This same warm layer is potentially the origin of other relevant spectral features such as the Balmer break, absorption, and the shape and relative intensities of the Balmer lines, as discussed in Sect.~\ref{sec:balmer_lines_origin} below.

\subsection{Insights from the Balmer lines}\label{sec:balmer_lines_origin}

The maximum \Hbeta{} and \Halpha{} emissivities in our models are reached close to the illuminated face of the cloud ($r/R_{\rm cloud}\lesssim 10^{-5}$), similar to the \ion{He}{i} lines (Fig.~\ref{fig:BHS_line_emissivities}). In this layer, the line emission is mainly driven by photoionzation and recombination radiation. However, due to the relatively small size of this layer, the contribution of this emissivity to the total can only be subdominant.
The ratio of the observed Balmer lines $\Halpha / \Hbeta = 10.4 \pm 0.3$ ($9.0\pm 0.8$ for the exponential wings only) and $\Hgamma / \Hbeta = 0.14 \pm 0.03$ ($0.13 \pm 0.04$) is far from the theoretical Case B recombination values of 2.86 and 0.47, respectively, for an optically thin, photoionized gas, or the value of $\Halpha/\Hbeta = 3.1$ that is often adopted for AGN narrow-line regions \citep{osterbrock2006}. An explanation for the high ratio could be that there is a significant contribution from collisional excitation in addition to photoionization to the fluxes of the Balmer lines \citep[e.g.,][]{raga2015}. Despite an ionized fraction of 0.99 in the hot layer, the neutral hydrogen density is still a few times $10^8$~cm$^{-3}$, making such collisional excitation scenario plausible.
In the high density, optically thick limit, the relative populations of the $n=3$ and $n=4$ levels tend to obey the Boltzmann distribution, resulting in collisionally excited lines with an \Hbeta{}/\Halpha{} that declines exponentially with decreasing temperature.
Following a similar analysis as presented in \citet{nikopoulos2025}, and using a standard \citet{cardelli1989} dust attenuation law, we obtain $\rm E(B-V) = 1.15 \pm 0.10$ from the \Halpha{}/\Hbeta{} broad exponential component ratio, and $\rm E(B-V) = 2.6 \pm 0.5$ from \Hgamma{}/\Hbeta{}. Such values would imply an extreme dust obscuration \citep[e.g.,][]{woodrum2025}. Moreover, the attenuation coefficients obtained from both Balmer ratios are in strong tension with each other, making it unlikely that dust attenuation is driving the observed Balmer decrements.

\begin{figure}
    \centering
    \includegraphics[width=\linewidth]{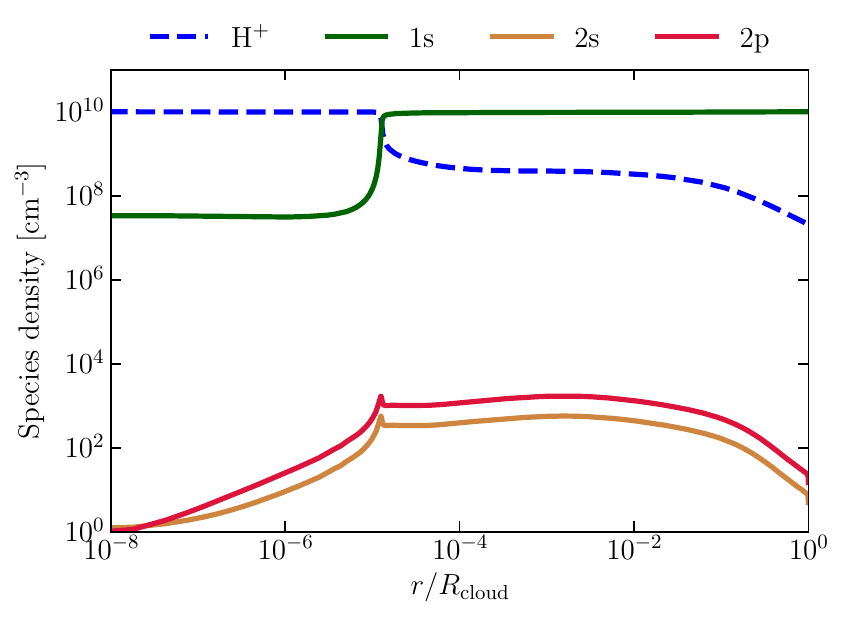}
    \caption{{\bf Density of hydrogen species inside a photoionized slab.} We show the volumetric density of fully ionized hydrogen (blue dashed) and first three levels (1s, 2s, 2p) of neutral hydrogen (solid lines) as a function of depth into the photoionized cloud of Fig.~\ref{fig:BHS_line_emissivities}. Although most neutral hydrogen in the envelope is in the 1s state, there is a significant population ($\eden\approx 10^3$~\unit{cm^{-3}}) of electrons in the 2s state, which produces the Balmer break, and resonant scattering of Balmer lines ($N({\rm H\,\textsc{i}, 2s}) \approx 10^{16}$~\unit{cm^{-2}}). The high column density of free electrons in the same region could be responsible for electron scattering broadening of the Balmer lines.}
    \label{fig:Hpop}
\end{figure}

As the photons travel outward, the high column density of gas in the 2s state in the thermalized layer will cause resonant scattering, which impacts H$\alpha$ and H$\beta$ photons differently. Resonant scattering mainly impacts the core profile of the line, where absorption in the partially ionized layer ($\eden/\Hden\sim 0.01$) causes the blue-shifted absorption profile \citep{Chang25}. In our Cloudy slab models, we find that the density of H atoms in the 2s state first increases sharply outward (by two orders of magnitude), peaking at the edge of the hot region, after which it remains relatively constant at a density of $n_{H, 2s} \sim10^3$ cm$^{-3}$ in the warm extended layer). In Fig.~\ref{fig:Hpop} we show the density of the fully ionized and three first levels of neutral hydrogen for a slab model with $\Hden = 10^{10}$~\unit{cm^{-3}} and $\NH = 10^{25}$~\unit{cm^{-2}} and our fiducial AGN SED with $\logU = -3$. A column density of $N({\rm H\,\textsc{i}, 2s}) \approx 10^{16}$~\unit{cm^{-2}} is reached in this model---a similar configuration produces electron scattering dominated \Halpha{} and \Hbeta{} with P Cygni cores in \citet{Chang25}. The gas layer can also explain the broadening of the Balmer lines by electron scattering. The nominal Thomson optical depth in this layer is $\tau_{\rm e}\sim 0.01$--0.1, slightly lower than what is predicted in \citet{Chang25} for an observed $FWHM\approx 1500$~\unit{km.s^{-1}} at $\sim 10^4$~K, but resonant scattering can greatly increase the path length of Balmer photons, hence the effective Thomson opacity, making this hypothesis viable.

The high opacity of this envelope gas to the Balmer transitions is seen as the Balmer absorption present in both \Halpha{} and \Hbeta{}. In a gas that is optically thick to Balmer lines due to the metastable 2s state, higher Balmer transitions are efficiently converted into \Halpha{}, pushing the \Halpha{}/\Hbeta{} ratio to even higher values \citep{Chang25}. This is an effect analogous to the Lyman Case B approximation \citep[the so-called Balmer on-the-spot approximation; e.g.,][]{osterbrock2006}. A combination of this effect with the aforementioned collisional excitation effects of high column density gas are likely driving the steep Balmer decrement observed in GN-9771 and other LRDs in the literature.

One striking feature is the fact that the Balmer absorption appears to be relatively stronger in \Hbeta{} than in \Halpha{}, whereas it is theoretically expected to be the opposite. This has previously also been observed in other LRDs \citep[e.g.,][]{deugenio2025_absorber}.
However, as discussed in Sect.~\ref{sec:balmer_O3}, our fiducial model also admits an absorption component with a fixed $\tau(\Hbeta)/\tau(\Halpha)=0.137$, albeit yielding a significantly worse BIC value ($\rm\Delta BIC \approx 16$). Therefore, our fiducial model still predicts a tension with the theoretical optical depth ratio, although the tension can be alleviated when taking into account the degeneracies of the emission components in the line core.
On the other hand, a tentative deviation of the theoretical $\tau(\Hbeta)/\tau(\Halpha)=0.137$ could be explained by different Balmer lines being effectively emitted at slightly different depths into the cloud. We note that the optical depth to Balmer absorptions in a neutral dense gas is very high, facilitating this effect within not very large physical distances. This hypothesis could also explain the tentative difference in the exponential wing widths of \Halpha{} and \Hbeta{} (see Sect.~\ref{sec:balmer_O3}), as both lines would potentially experience different electron column densities. Alternatively, this result could be attributed to differences in the absorption infilling, if relatively higher fraction of \Halpha{} emission is emitted at the edge of the cloud compared to H$\beta$ (see Fig.~\ref{fig:BHS_line_emissivities}). This could occur in the case that \Halpha{} and \Hbeta{} emission is driven by collisions in the edge of the warm layer, also evidenced by the lack of P Cygni emission in \Hbeta{}. The P Cygni components of \Halpha{} and \Hbeta{} (Fig. $\ref{fig:Ha_line_fit}$) interestingly have comparable line widths as the [\feii{}] emission, suggesting that both the Balmer lines and [\feii{}] could indeed partially originate from the warm layer. Alternatively, this component could originate from the resonantly scattered H$\beta$ photons that branched to H$\alpha$.
We emphasize that due to the strong degeneracies of our fiducial model used to fit \Halpha{} and \Hbeta{}, addressing the flux and spectral shape of this infilling component remains challenging (see Sect.~\ref{sec:balmer_O3}). Generally, more tailored radiative transfer simulations \citep[e.g., building upon the model presented in][]{Chang25}, incorporating the effects of collisional excitation, are required to further study the exact conditions yielding the high H$\alpha$/H$\beta$ ratios and explaining the detailed differences in the line-profiles.

Contrary to \Halpha{} and \Hbeta{}, the higher Balmer transition line \Hgamma{} seemingly contains an important contribution of a narrow component, compatible with the width of \oiii{} $\lambda\lambda 4960, 5008$. This result is in line with the interpretation that this narrow component is produced by a host galaxy. Under this assumption, the narrow components of all the Balmer lines would follow line ratios closer to the typical Case B ratios, assuming a low dust attenuation. Therefore, the narrow \Hgamma{} emerges, whilst in \Hbeta{} and \Halpha{} the respective narrow components are outshined by the broad, collisionally  and/or scattering-dominated emission with higher relative intensities. In this scenario, the narrow \Hgamma{} component is a compelling tracer of the host of GN-9771, which we discuss in Sect.~\ref{sec:host_implications} in detail.

\subsection{Insights from Helium emission lines} \label{sec:hei_origin}

Unlike as is the case for the Balmer lines, the dense gas is optically thin to other lines, for example  \ion{He}{i}, \ion{He}{ii}, and [\feii{}], making these promising probes of the internal dynamics and ionization conditions of LRD envelopes \citep[e.g.,][]{BWang25}, since they traverse it without being heavily affected by self-absorption (except perhaps by electron scattering). Although the HeI 4472, 5877, 7067 lines are much fainter than the Balmer lines and therefore detected only with modest signal to noise, they appear to show broad wings (Section $\ref{sec:HeI}$, see Fig. $\ref{fig:feii_fit_1}$). This is in line with our modeling that suggest they originate in the same gas as the Balmer photons. As the cross-section to Thomson scattering is independent of wavelength, this will yield similar broad wing, but they are not subject to resonant scattering such that the photons in the line-center escape through the warm layer more easily. 

The \ion{He}{i} $\lambda$5876 line is a strong recombination line, often found in the spectra of Type 1 AGN with broad profiles \citep[e.g.,][]{vandenberk2001, kuhn2024}. This line arises from de-excitation to the triplet level 2$^3$S of He from higher levels, along with other triplet lines (\ion{He}{i} $\lambda$4472, 7067). However, 2$^3$S is a metastable level, therefore in dense media it can be depopulated via collisions, in particular in absence of a strong ionization field \citep{mathis1957, benjamin2002}. The observed ratios of the \ion{He}{i} lines are therefore highly sensitive to density and temperature, in particular for high gas column densities \citep{almog1989}.

Particularly interesting are the observed \ion{He}{i} singlet lines, which intensity can be enhanced at very high densities if collisional effects are taken into account \citep[e.g.,][]{benjamin2002}.
These lines are rarely reported in spectra of most AGN populations \citep[e.g.,][]{veron2002}. GN-9771 presents unusually high flux ratios of these lines when compared to the triplet line \ion{He}{i} $\lambda$5876: $\lambda5876/\lambda5017 = 1.18 \pm 0.4$. Using the Python tool {\sc Pyneb} \citep{luridiana2015}, we estimate this ratio to be $\sim$6 in an optically thin, purely recombination driven scenario, for a broad range of gas density and temperatures. We note, however, that the \ion{He}{i} $\lambda$5017 line is potentially blended with \oiii{} and [\feii{}], and the measured flux could be affected by these other lines.
The \ion{He}{i} singlet lines are also found to be strong in Type IIn supernovae, which are believed to have dense opaque gas envelopes \citep{groeningsson2007,dessart2009} and therefore present similar configurations as those being considered for LRDs \citep[e.g.,][]{ji2025,naidu2025}.

The strongest \ion{He}{i} line in the observed wavelength range is $\lambda$7067. The ratio \ion{He}{i} $\lambda$5876/$\lambda$7067~$=0.85$ is low in comparison with the typical values of 1.8--3 in star-forming galaxies at similar redshifts \citep[e.g.,][]{yanagisawa2024}. The $\lambda$7067 line is strongly affected by complex radiative transfer and density effects, particularly for high densities, that could enhance its emission through pumping from the 2$^3$S level \citep[e.g.,][]{benjamin2002, berg2025}. Our photoionized cloud models fail to reproduce the high \ion{He}{i} $\lambda$5876/$\lambda$7067 ratio (see Fig.~\ref{fig:BHS_line_emissivities}). More complex models that take these effects into account are needed in order to investigate \ion{He}{i} ratios and intensities as density and temperature tracers, but generally the strengths of the singlets and the $\lambda$7067 line is a further proof that dense gas constitutes a central component of LRDs emission.
The \ion{He}{ii} $\lambda$4687 line is not detected in GN-9771; in line with the low emissivity predicted by our models (integrated emissivity ratio \ion{He}{ii}$\lambda$4687/\ion{He}{i}$\lambda$5876 = 0.05; see Fig.~\ref{fig:BHS_line_emissivities}).

\begin{table*}
\centering
\caption{Summary of the observed features in the FRESCO-GN-9771 LRD and their interpretation in the context of the black hole star photoionization model. }

\begin{tabular}{p{3.4cm}|p{5cm}|p{4cm}|p{3.7cm}}
\hline
\textbf{Feature} & \textbf{Interpretation in the BH* model}  & \textbf{Alternatives/Challenges} & \textbf{Implication} \\ \hline
Broad wings in Balmer lines &  Lines broadened by electron scattering in the warm layer with $N(\text{\ion{H}{i}}, \rm 2s) \approx 10^{16}$~\unit{cm^{-2}}. & Alternatively would mainly trace motions in broad-line region & Virial BH mass scaling relations not applicable.  \\ \hline

\makecell{$\Halpha / \Hbeta = 10.4 \pm 0.3$ \\ $(\gg 2.8)$} & Indicates collisional excitation,
but could also partly be impacted by resonant H$\beta$ scattering, both indicative of very high densities.
& Alternatively would imply very high dust attenuation at odds with continuum slope. & Dust corrections based on Balmer decrement are not reliable. Bolometric corrections need revision. \\ \hline

P Cygni features in \Halpha{} \& \Hbeta{} & High opacity from layer of warm partially ionized outflowing gas. & \Hbeta{} absorption may be unexpectedly stronger  than \Halpha{}, but could be \Halpha{} infilling, or emission at different physical depths.  & The outer layer is outflowing$^\dag$. \\ \hline

Shape and strength of Balmer break & High opacity from the turbulent, dense gas in the warm outer layer. & Alternatively, it would require highly unusual stellar populations with almost exclusively A stars. & Optical continuum does not originate from stellar light from the host galaxy, i.e., uncertain stellar masses. \\ \hline

Optical [\feii{}] forest & Arises from a warm and partially ionized layer with $\Te = 6000$--7500~K. & Photoionized by AGN in an intermediate-density region & No strong photoionization needed. Broader lines than \oiii{} point toward different origin (not from a narrow-line region). \\ \hline

UV \feii{} & Observed absorption troughs compatible with main resonant \feii{} multiplets. & Strong UV \feii{} emission is possible, but difficult to concile standard templates and models. Overall shape of the UV is very uncertain and different from typical SFGs and AGN. & Deep high-resolution data needed to resolve the UV features. \\ \hline

Very narrow \oiii{} $\lambda\lambda$4960,5008 and H$\gamma$ emission & Emission from \ion{H}{ii} regions around young stars of a typical low-mass host galaxy & No other such narrow lines are convincingly detected. Young stellar cluster associated with the BH?  & Suggests that the host galaxy is low mass (M$_{\rm dyn}\sim3\times10^9$ M$_{\odot}$, M$_{\rm star} \sim 10^8$ M$_{\odot}$)  \\ \hline

BH mass & $L_{\rm bol}$ suggests $M_{\rm BH}/M_\odot \sim 0.1$ assuming Eddington-rate accretion. Could be even lower if higher accretion rates apply. & Applying standard scaling relations based on \Halpha{} would imply $M_{\rm BH}/M_\odot > 1$. &  $M_{\rm BH}/M_\odot$ tension with local relation can be greatly alleviated with reasonable (super-)Eddington ratios. \\
\hline

Low \ion{He}{i} $\lambda$5876/$\lambda$7067 ratio  & The model cannot reproduce such relatively strong \ion{He}{i}$\lambda$7067 emission. & Pumping contribution and density effects may boost the relative luminosity of $\lambda$7067. Better RT models needed. & Showcases Helium lines as an additional probe of the temperature and density structure. \\ \hline

\end{tabular}

\tablefoot{
$^\dag$Some LRDs in the literature present redshifted P Cygni, and thus they could be interpreted as inflows. Some show different absorption velocities for different Balmer transitions, indicating complex velocity and opacity effects.
}

\label{tab:emission_features}
\end{table*}

\section{Implications} \label{sec:implications}

\subsection{Implications for the galaxies hosting LRDs}\label{sec:host_implications}

Given that GN-9771 appears to be a point-source at all observed wavelengths and that a significant fraction of the emission can be explained by the AGN emission in the context of the BH* model (Fig. $\ref{fig:prism_w_cloudy_BHS}$), an important question to ask is where and what the host galaxy of GN-9771 is. Characterizing host galaxy is important to  better understand the context of BH formation models and SMBH-galaxy co-evolution.

Here we attempt to quantify the fraction of the UV continuum that originates from the host galaxy. We assume that the key feature that traces the host galaxy emission is the \oiii{} emission due to its extremely narrow line width, even compared to AGN-associated lines as [\feii{}]. The \oiii{} width implies a dynamical mass of $\logten (M_{\rm dyn} / M_\odot) = 9.6 \pm 0.5$, using the prescriptions from \citet{bezanson2018}. For typical galaxies, such dynamical mass would correspond to a stellar mass of $\logten(M_*/M_\odot) \approx 8$, applying the scalings found in \citet{degraaff2024_Mdyn}, or $\logten(M_*/M_\odot) \approx {8.6}$ according to \citet{saldana-lopez2025}. This is a very similar stellar mass was found by \citet{deugenio2025_absorber} for an LRD at a similar redshift and in line with clustering results \citep{matthee2024b}. Following this result, we can test the maximum contribution to the UV light from a blue host galaxy. We measure the UV absolute magnitude of GN-9771 to be $M_{\rm UV} = -19.3 \pm 0.4$, based on the median flux density in the rest-frame 1400--1600~\AA{} interval. For comparison, the median UV magnitude of 38 galaxies at $5<z<6$, with masses $8<\logten(M_*/M_\odot)<8.6$ in the catalog of the ALT survey \citepalias{naidu2024} is $M_{\rm UV} = -18.7$ (ranging from $-20.3$ to $-17.3$). These results are therefore consistent with some contribution of a host galaxy to the UV spectrum of GN-9771, with the measured \oiii{}, within the uncertainties given the broad scatter in $M_{\rm dyn}/M_*$. In Fig.~\ref{fig:prism_w_cloudy_BHS} we include the SED fit by prospector to the object ALT-31334 ($z=5.66$) as a typical example: a galaxy with $\logten(M_*/M_\odot) = 8.0$ and $M_{\rm UV}= -18.7$, as expected for the host galaxy of GN-9771. It has a typical \oiii{} equivalent width of $\approx700$~\AA{}; for comparison, the\oiii{} $\lambda\lambda 4960,5008$ EW of GN-9771 is $370 \pm 14$~\AA{} \citep[see, e.g.,][]{Matthee23,Endsley24}. This example illustrates how a typical low-mass galaxy can explain the observed narrow \oiii{}, a minor fraction of the narrow Balmer emission-lines and with a minor contribution to the overall UV-optical spectrum of the LRD, which is only dominant at $\lambda_0<1500$ {\AA}.

A surprising aspect of GN-9771's UV spectrum is its noteworthy similarity with the luminous LRD A2744-45924, with also a very similar $M_{\rm UV} = -19.2$. In particular, they share the same shape around 2400--3000~\AA{}, which we argued in Sect.~\ref{sec:uv_feii} it can be partly shaped by low-ionization \feii{} absorption. Identical UV features are also present in other luminous LRDs \citep[e.g.,][]{tripodi_deep_2025, deugenio2025_irony}. Such features are characteristic of low-ionization absorption systems, frequently observed in FeLoBAL quasars \citep[e.g.,][]{zhang2022}. However, under the assumption that the 2500~\AA{} bump is a result of resonant UV \feii{} multiplet absorptions, the rest of the UV features are hard to explain. It is possible that this part of the SED of GN-9771 is shaped by multiple overlapping absorption troughs, which is not unusual in FeLoBAL systems, as their shapes is very diverse \citep[e.g.,][]{hall2002}. Notably some FeLoBALs show similar Balmer absorptions as reported in LRDs \citep{leighly2025}.
The strong similarity of the UV spectrum of GN-9771 and other luminous LRDs \citep[][]{tripodi_deep_2025, labbe2024, deugenio2025_irony} suggests that these objects must have significant contribution from AGN to their UV light, as the observed features are hard to explain with a dominating star-forming host. This result is in line with the spatially resolved offset found in \citet{torralba2025} between the compact \Halpha{} emission and the far-UV and \lya{} of A2744-45924, the latter associated with the host galaxy.
Deeper and high-resolution data would be needed to investigate the UV features of GN-9771 in detail, since the low PRISM resolution difficultates this task.

Additionally, the narrow components of the Balmer lines (see Sect.~\ref{sec:balmer_O3}) also provide insights into the properties of an hypothetical host galaxy. The modelling of the \Hgamma{} line is challenging due to blending with \oiii{} $\lambda 4364$ and multiple strong [\feii{}] lines in the vicinity. Assuming a saturated \Hgamma{} absorption by the dense gas---that is, the absorption only affects the broad component of this line, and not the continuum---we obtain a narrow \Hgamma{} flux of $0.6\times10^{-18}$~\unit{erg.s^{-1}}. This flux can be regarded as a lower limit in the case that the \Hgamma{} absorption would significantly affect the continuum.
Assuming Case B ratios, and disregarding dust attenuation in the host galaxy, this flux would correspond to a star formation rate of $\sim 5$~\unit{M_\odot.yr^{-1}}, using the SFR--\Halpha{} calibration from \citet{kramarenko2025}. Such SFR typically corresponds to star-forming galaxies at $z\approx 5$ with $\log_{10} ( M_* / M_\odot) \approx 7.7$--9.3\footnote{We remark that here we assume no AGN contribution to the narrow \Hgamma{} flux.} \citep[e.g.,][]{dicesare2025}, consistent with our estimate based on the dynamical mass. Still following the assumption of narrow Case B ratios, we obtain $\oiii{} \lambda 5008 / \Hbeta_{\rm Narrow}\approx 5$, which implies a gas-phase metallicity of $12+\log_{10}(\rm O/H)\sim 7.4$, slightly toward the low end of the mass-metallicity relation, but within the scatter \citep[e.g.,][]{chakraborty2025}.

\subsection{The nature of LRDs}\label{sec:LRD_nature}

In Sect.~\ref{sec:balmer_lines_origin} we have presented multiple insights into the Balmer lines in the context of a dense gas envelope, summarized in Table~\ref{tab:emission_features}.
The exponential wings in the Balmer lines indicate a significant contribution to line broadening beyond gas dynamics \citep[e.g.,][]{naidu2025, rusakov2025, Chang25}, unlike what is typically assumed for the quasar BLR. 
The P Cygni shape of the \Halpha{} and \Hbeta{} line cores result as a consequence of high opacity and resonant scattering in the warm envelope \citep[e.g.,][]{Chang25}. Therefore, the shape of the P Cygni responds to the gas dynamics of the edge of the cloud, which in the case of GN-9771 is outflowing but could in other cases be inflowing (such as A2744-45924; \citealt{labbe2024}). These results imply that the mechanisms governing the shape of \Halpha{} and \Hbeta{}---both in the line core and the broad wings---are very different from typical BLR conditions, as discussed in detail in Sect.~\ref{sec:balmer_lines_origin}. Due to the multiple differences in the Balmer lines, the standard scaling relations used to infer black hole masses and bolometric luminosities in quasars should, in general, not be applicable to LRDs.

The optical spectrum of GN-9771 does not show strong high-ionization lines, such as \ion{He}{ii} $\lambda$4687, suggesting a cut-off in the incident spectrum (just) below $\approx50$ eV \citep[e.g.,][]{BWang25}. It also does not show lines with a low critical density, such as [\ion{N}{ii}] or [\ion{S}{ii}]. However, the spectrum does show \oiii{} $\lambda$4364 line.
GN-9771 neither shows high-ionization lines in the UV (e.g., \ion{C}{iv} $\lambda 1549$, \ion{C}{iii} $\lambda 1909$, \ion{Mg}{ii} $\lambda 2799$, typically strong in Type I \& II AGN).
In turn, collisional excitation seems to be an important mechanism for at least some of the optical lines, like the Balmer lines and [\feii{}] \citep[see, e.g.,][]{kwan1981, shields2010, sameshima2011}. Moreover, the high \Halpha{}/\Hbeta{} ratio may suggest a very significant contribution of collisional excitation processes to these lines (see Sect.~\ref{sec:balmer_lines_origin} for a detailed discussion). An alternative explanation for the Balmer decrement would be very high dust attenuation, but this is at odds with, for example, the relatively blue optical colors. We note that broadband LRD spectra show significant variation (Fig. $\ref{fig:prism_spectrum}$), especially in their UV spectrum. Whether such variations could be attributed to variations in host galaxy emission or in AGN conditions will be explored in a future work.

\begin{figure}
    \centering
    \includegraphics[width=\linewidth]{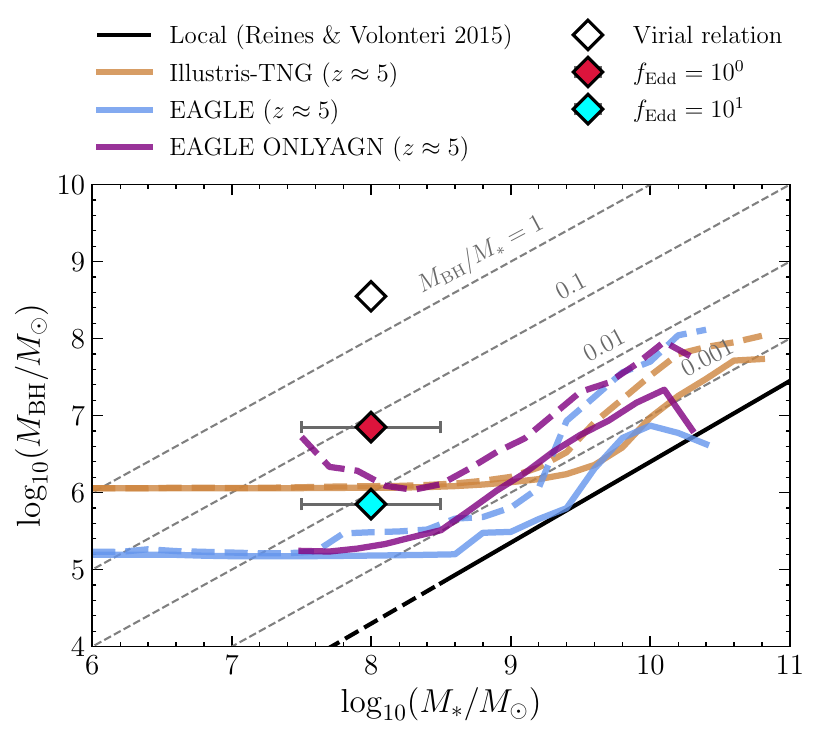}
    \caption{{\bf BH vs. Stellar mass relation for GN-9771.} We show estimations of $M_{\rm BH}/M_*$ for GN-9771 based on a stellar mass inferred from the dynamical mass suggested by the width of \oiii{}, and $M_{\rm BH}$ estimated using a standard virial relation (empty square; \citealp{matthee2024a}), and assuming an Eddington luminosity ratio of $f_{\rm Edd}=1$ (red square), and $f_{\rm Edd}=10$ (cyan square). We make a comparison with the local relation \citep{reines2015} and the Illustris-TNG \citep{weinberger2017, pillepich2018} and EAGLE \citep{schaye2015, crain2015} simulations at $z\approx5$. We highlight the median trend with a solid (dashed) line and the extrema (95th percentiles) of the distributions in each simulation.}
    \label{fig:Mstar_MBH}
\end{figure}

Using standard local virial calibrations, \citet{matthee2024a} obtained a BH mass of $M_{\rm BH} = 10^{8.55}$~$M_\odot$ for GN-9771 \citep{greene2005}, fitting a broad Gaussian for a broad-line region. We estimate the bolometric luminosity of GN-9771 by integrating the best-fitting Cloudy slab model for $\lambda > 912$~\AA{} \citep[motivated by][]{greene2025}, obtaining $L_{\rm bol} = 10^{44.95}$~\unit{erg.s^{-1}}. This bolometric luminosity would imply a very low Eddington ratio of $L_{\rm bol} / L_{\rm Edd} = 0.02$.
This low value is in contrast with the super-Eddington scenarios recently proposed to explain many features of LRDs such as low continuum variability \citep[e.g.,][]{secunda2025, furtak2025} or the lack of X-ray detections \citep[e.g.,][but see also \citealp{sacchi2025}]{inayoshi2024, madau2025, madau2024b}.

Recent works have proposed models that describe many observed features of LRDs as photospheric emission from BH envelopes \citep[e.g.,]{kido2025, begelman2025, zwick2025, liu2025}. In particular, \citet{begelman2025} proposed a late-quasi-star scenario that describes many observed features of LRDs. The osberved properties of GN-9771 are very compatible with this scheme, which outlines a BH of mass $M_{\rm BH}\sim 10^6\ M_\odot$ accreting at super-Eddington rates, yielding a luminosity of $\sim 10^{44\text{--}45}$~\unit{erg.s^{-1}}, with optical color temperatures of $\sim6000$--7000~K (similar to the temperature of our modeled [\feii{}] emitting envelope). Similarly, \citet{liu2025}, using semi-analytical atmosphere models, found that a super-Eddington system could yield an envelope with effective temperatures of $4000$--$6000$~K that effectively reproduce the Balmer break and red optical colors of LRD, favoring models with high (super{-})Eddington accretion rates.
If Eddington-luminosity rate is assumed ($f_{\rm Edd} \equiv L_{\rm bol} / L_{\rm Edd} = 1$), our measured $L_{\rm bol}$ would imply $M_{\rm BH} \approx 10^{6.85}$~$M_\odot$, which would correspond to an upper limit if super-Eddington accretion is invoked \citep[see also][]{kokorev2023, wang2025}. In combination with our stellar mass estimate \citep[using the more conservative calibration of][]{degraaff2024_Mdyn}, this result yields $M_{\rm BH}/M_* \approx 0.1$. If a higher $f_{\rm Edd}=10$ is assumed, consistent with the value of $\dot{M}/\dot{M}_{\rm Edd}=10^2$ favored by the models of \citealp{liu2025} \citep[see][]{inayoshi2020}, we obtain a BH mass of $M_{\rm BH} \sim 10^{5.85}\ M_\odot$, yielding $M_{\rm BH}/M_* \approx 0.01$, greatly alleviating the tension with the local relation \citep[][]{reines2015}.
In Fig.~\ref{fig:Mstar_MBH} we show our different mass estimates in comparison with the observed local relation and simulated galaxies in the Illustris-TNG \citep{weinberger2017, pillepich2018} and EAGLE \citep{schaye2015, crain2015} simulations at $z\approx 5$. These simulations have been calibrated to match the BH to stellar mass ratio in massive galaxies at $z\sim0$, but have not been tested in the early Universe. In the standard EAGLE and Illustris-TNG simulations, we notice that SMBHs in galaxies with masses $\sim10^8$ M$_{\odot}$ are almost exclusively at the seed mass, i.e., there is no gas fueling SMBH growth yet, clearly at odds with observations of AGN in such low-mass galaxies, regardless of what the SMBH masses are. Interestingly, the EAGLE version without stellar feedback (ONLYAGN) enables SMBH accretion in lower mass galaxies, with some of the extremes having SMBHs with masses a few times $10^6$ M$_{\odot}$. This indicates that simulations of early galaxies with higher resolution that enables more efficient cooling of the ISM and SMBH growth \citep[e.g.,][]{Chaikin25} may contain galaxies hosting LRD phenomena.

\subsection{Comparison to {\rm [\feii{}]} in other LRDs}\label{sec:other_lrds_stacks}

Thanks to our new deep spectrum of GN-9771, we identify numerous previously poorly studied features of LRDs. Many of these emission-lines have been historically studied in the context of AGN, although their signatures differ in details. For example, while broad \feii{} emission is commonly studied in quasar spectra, the relatively narrow forbidden [\feii{}] emission is rarely seen in quasars. One notable exception is the quasar SDSS J1028+4500 at $z=0.58$ \citep{wang2008}. The spectrum of this object shows a very similar [\feii{}] complex to that of GN-9771. Remarkably, it also shows significant \Hbeta{} and \ion{He}{i} absorption, which is very rare among quasars. As we discussed in Sect.~\ref{sec:emline_origin}, the narrow [\feii{}] lines are likely emitted in an intermediate-density warm gas with a high column density ($\NH \gtrsim 10^{24}$~\unit{cm^{-2}}). This same gas is arguably also responsible for the Balmer absorption seen in both \citet{wang2008} and GN-9771---and, by extension, in other LRDs. 
Recently, [\feii{}] has been clearly detected in a local ($z=0.1$) LRD \citep{lin2025, ji2025_LoL}, the $z=2.3$ \textit{Rosetta Stone} \citep[][see also \citealp{juodzbalis2024}]{ji2025_LoL}, and the LRD at $z=6.7$  RUBIES-EGS-49140 \citep{lambrides2025, deugenio2025_irony}.
The question remains if [\feii{}] can be ubiquitously detected in LRDs with sufficiently deep data.

\begin{figure}
    \centering
    \includegraphics[width=\linewidth]{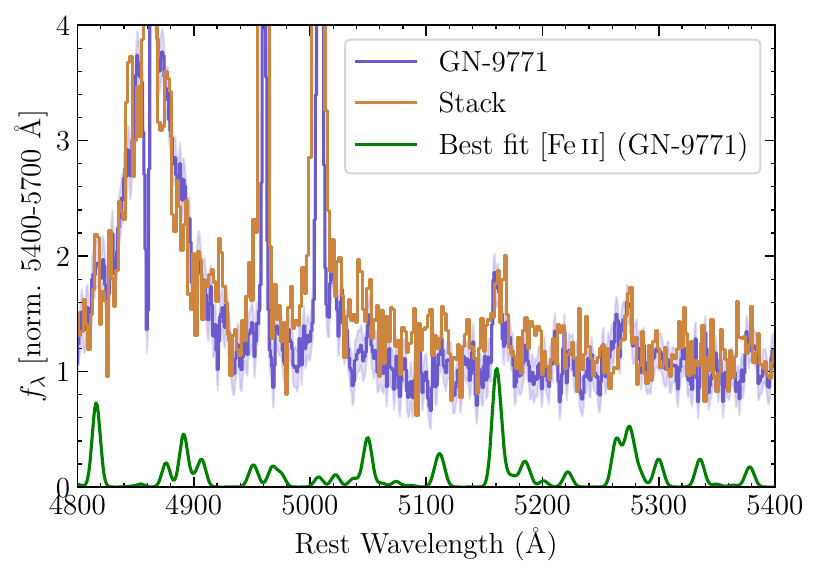}
    \caption{{\bf Comparable [\feii{}] emission in a stack of LRDs}. We compare the [\feii{}] optical spectrum of GN-9771 with a stack of 13 medium-resolution grating spectra of LRDs. The spectrum of GN-9771 is degraded to the medium resolution of the NIRSpec M-gratings for this comparison. The most luminous [\feii{}] lines are detected in the stack, suggesting that these are common among LRDs, although their weak intrinsic fluxes challenge their detection.}
    \label{fig:stack_feii_all}
\end{figure}

In order to investigate the prevalence of narrow [\feii{}] emission in high-redshift LRDs, we stack publicly available medium- and high-resolution grating spectra of a sample of 13 LRDs at $z=4.13-6.98$, covering H$\beta$ and H$\alpha$ to attempt detecting the [\feii{}] lines also observed in GN-9771. In Fig.~\ref{fig:stack_feii_all} we compare the grating spectrum of GN-9771 and the fit [\feii{}] to a stacked sample of four objects in our NIRSpec IFU program (excluding GN-9771 itself; see Sect.~\ref{sec:data})\footnote{FRESCO-GN-12839, FRESCO-GN-15498, FRESCO-GN-16813 and FRESCO-GS-13971 from \cite{matthee2024a}.} plus nine medium-resolution spectra of LRDs from the JADES \citep{eisenstein2023} and RUBIES \citep{degraaff2025_rubies} surveys, publicly available in the Dawn JWST Archive.\footnote{\url{https://dawn-cph.github.io/dja/}; Specifically, JADES-GN-38147, JADES-GN-68797, JADES-GN-73488, RUBIES-EGS-42046, RUBIES-EGS-49140, RUBIES-EGS-50052, RUBIES-EGS-55604, RUBIES-UDS-182791 and RUBIES-UDS-47509.} For stacking, the high-resolution G395H spectra were degraded to match the resolution of the M grating ($R\sim 1000$). We align each spectrum to a common rest-frame wavelength grid and normalize each spectrum to the average continuum flux density over 5300-5500 {\AA} before we obtain a median stack. In the stacked spectrum, the most prominent [\feii{}] features are indeed detected ($\lambda$4816, 5160, and 5275) with line ratios comparable to those seen in GN-9771.

The variation among the strength  of the optical [\feii{}] emission, and the relation with other LRD properties (e.g., strength of the Balmer break, emission line properties etc.) should be tested with very deep grating spectroscopy in the future, due to the low intrinsic EW of [\feii{}]. Assessing the prevalence of [\feii{}] emission in LRDs will determine how our results apply to such broad population.

\section{Summary}\label{sec:summary}

In this work, we  employed JWST/NIRSpec IFU PRISM and high-resolution G395H data to investigate the properties of the luminous LRD FRESCO-GN-9771 at $z=5.535$, and interpret them in the context of dense-gas envelope models (BH* models). In Table~\ref{tab:emission_features} we outline our interpretation of multiple common LRD spectral features in the context of BH* models. Below we summarize our main findings in this paper.

\begin{enumerate}

    \item GN-9771's \Halpha{} and \Hbeta{} emission line profiles are well described by a model consisting of broad exponential wings up to at least $\pm \sim 7000$~\unit{km.s^{-1}}, and a P Cygni profile in the line cores ($\pm\sim 1000$~\unit{km.s^{-1}}). The exponential wings are in line with broadening by Thomson scattering, while the P Cygni cores may originate in a dense gas layer with high opacity to Balmer transitions. A broad \Hgamma{} is also detected, albeit with a lower S/N (Sect.~\ref{sec:balmer_O3}, Figs.~\ref{fig:Ha_line_fit} and~\ref{fig:Hb_line_fit}).

    \item We detected a forest of multiple [\feii{}] transitions in the rest-optical spectrum of GN-9771. These transitions are fit with fixed theoretical ratios. Their relative strength is high ($[\feii{}]/\Hbeta\approx 0.13$ in the range 5100--5400~\AA{}), and they substantially contaminate the \Hbeta{} line profile, implicating that \feii{} should be taken into account for the line fitting of this line (Sect.~\ref{sec:forb_feii_model}).
    
    \item We measured a low \ion{He}{i} $\lambda5876/\lambda7067 \sim 0.81$ ratio, which is another indicator of high-density gas, and suggests that complex radiative transfer effects play a crucial role in the emission of these lines. We also detected the typically faint singlet lines \ion{He}{i} $\lambda$5017 and $\lambda$6680. (Sect.~\ref{sec:HeI} and~\ref{sec:hei_origin}).

    \item We tested photoionization models computed with Cloudy, to investigate the origin of GN-9771's emission lines. A BH*-like setup yields a best fit to the [\feii{}] ratios for a gas density of $\Hden= 10^\text{9--10}$~\unit{cm^{-3}} and column densities $\NH > 10^{24}$~\unit{cm^{-2}}, while still producing a Balmer break with a  strength that is similar to what is observed. The [\feii{}] emission is generated in a warm $\Te\approx 6000$--7500~K outer layer, shielded from strong ionizing radiation (Sect.~\ref{sec:phot_models}).

    \item The warm outer layer predicted by Cloudy models not only explains the [\feii{}] emission, but is also the origin of the strong Balmer break and Balmer P Cygni profiles due to absorption by dense neutral hydrogen gas with a significant column density where the $n=2$ state is significantly populated ($N(\ion{H}{i}, \rm 2s)\approx 10^{16}$~\unit{cm^{-2}}). Additionally, the unusually high $\Halpha{}/\Hbeta{}=10.4\pm3$ and $\Hbeta{}/\Hgamma{} = 7.1 \pm 1.5$ ratios are indicative that collisional effects and resonant scattering are significantly affecting the Balmer line emission. These processes happen in the warm, and partially ionized layer, which contains mostly neutral hydrogen, but also a fairly high electron density ($\eden / \Hden \sim 0.01$; Figs.~\ref{fig:BHS_Te_ne}, \ref{fig:BHS_line_emissivities}, and~\ref{fig:Hpop}).

    \item In addition to a similar broad wing, the \Hgamma{} profile  shows a narrow component that is virtually unseen in \Halpha{} or \Hbeta{}. We associate this narrow emission with the host galaxy, and argue that this component is subdominant in the other two Balmer lines due to outshining by the broader emission, dominated by collisional processes and resonant scattering with flux ratios that strongly deviate from Case B. Meanwhile, the narrow component of all the Balmer lines should have ratios more similar to the expected in Case B recombination and trace host galaxy light (Sect.~\ref{sec:host_implications}).

    \item The narrow \oiii{} and \Hgamma{} emission suggest a host galaxy stellar mass of $\sim 10^8$~\unit{M_\odot} and a star formation rate of $\sim 5$~\unit{M_\odot.yr^{-1}}. On the other hand, the bolometric luminosity of GN-9771, $\log_{10} ( L_{\rm bol} / \rm erg\,s^{-1})\approx 44.95$, suggests a black hole mass of $M_{\rm BH}\approx 10^{6.85}$~\unit{M_\odot} assuming an Eddington-luminosity ratio of $L_{\rm bol} / L_{\rm Edd} = 1$. This is roughly two orders of magnitude lower than local virial calibrations suggest. If higher Eddington ratios are assumed, the $M_{\rm BH} / M_*$ tension with the local relation and simulations is greatly alleviated (e.g., $L_{\rm bol} / L_{\rm Edd} = 10$ would imply $M_{\rm BH} / M_* \approx 0.01$; Sects.~\ref{sec:host_implications} and~\ref{sec:LRD_nature}).

\end{enumerate}

These results show that LRDs are composite sources of a low-mass star-forming host galaxy, with a central AGN source that dominates the optical and even a significant fraction of the UV light (especially beyond $\lambda_0>1500$ {\AA}). The AGN is embedded in dense yet metal-enriched gas that is primarily found in a thermalized layer with a temperature of $\sim7000$ K. The emerging opacity to Balmer photons explains the strong and smooth Balmer break in the AGN spectrum, whereas collisional excitation and scattering processes in the warm layer explain the Balmer emission lines and their line profiles. Whereas scattering processes lead to non-dynamical broadening of the Balmer lines, the sheer luminosity ($\sim10^{45}$ erg s$^{-1}$) and high gas densities remain the strongest evidence that an AGN is ultimately powering the emission. Nevertheless, many questions remain regarding  the origin of the circumnuclear gas and the stability of the envelopes.  Given that virial BH mass tracers may have become unreliable, perhaps  we can only approximate black hole masses assuming Eddington accretion, although we remain unsure whether these results are applicable to the whole population of LRDs. These and other questions will occupy our focus for years to come.

\begin{acknowledgements}

We thank the scientific referee for useful and constructive comments.

We thank Ylva Götberg and Zoltan Haiman for insightful discussions about the physics of gaseous envelopes and accretion into black holes.

Funded by the European Union (ERC, AGENTS,  101076224). Views and opinions expressed are however those of the author(s) only and do not necessarily reflect those of the European Union or the European Research Council. Neither the European Union nor the granting authority can be held responsible for them. 

This work is based in part on observations made with the NASA/ESA/CSA James Webb Space Telescope. The data were obtained from the Mikulski Archive for Space Telescopes at the Space Telescope Science Institute, which is operated by the Association of Universities for Research in Astronomy, Inc., under NASA contract NAS 5-03127 for JWST. These observations are associated with program \#5664.%

This work has received funding from the Swiss State Secretariat for  Education, Research and Innovation (SERI) under contract number  MB22.00072, as well as from the Swiss National Science Foundation (SNSF) through project grant 200020\_207349.

\end{acknowledgements}

\bibliographystyle{aa}
\bibliography{my_bibliography}

\appendix

\section{Continuum subtraction}\label{sec:continuum_subtraction}

\begin{figure*}
    \centering
    \includegraphics[width=\linewidth]{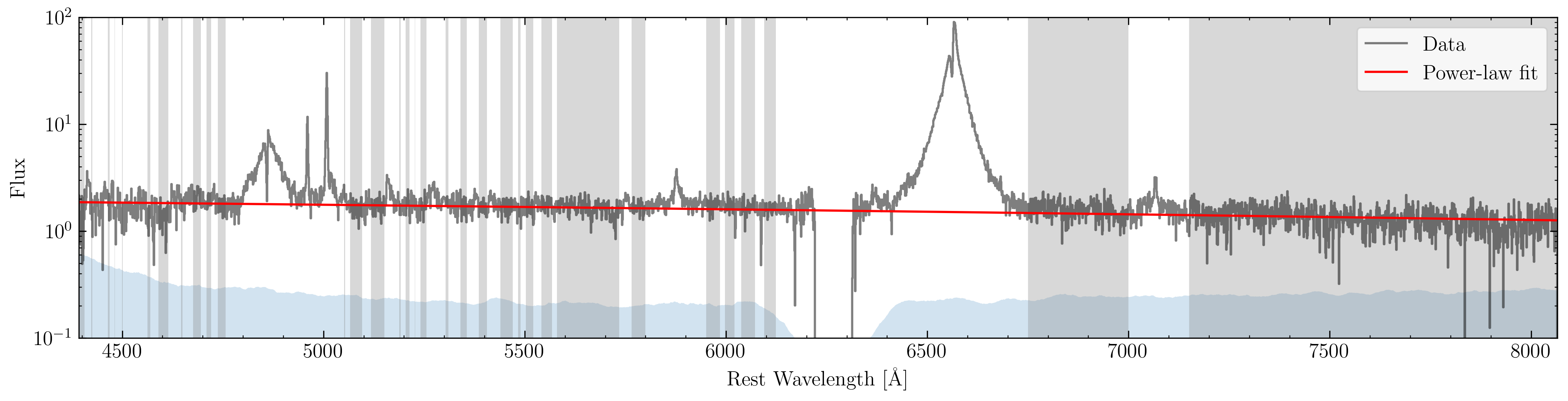}
    \caption{Power-law fit to the optical continuum of GN-9771, used for the \oiii{} and Balmer line fitting. We fit a power law to the optical $f_\lambda$ G395H spectrum, using the wavelengths shown as gray shaded regions. Between rest-frame $\sim$6200 and $\sim$6300 the lack of data corresponds to the NIRSpec detector gap.}
    \label{fig:continuum_subtraction}
\end{figure*}

In Fig.~\ref{fig:continuum_subtraction} we show the simple power-law fit to the continuum in the gray shaded ranges. This fit is done prior to the Balmer line fitting described in Sect.~\ref{sec:balmer_O3}, although in the fitting an extra component for the continuum level is further added. The continuum-fit mask is obtained by excluding the wavelengths affected by \Hbeta{}, \Halpha{}, \oiii{}, the multiple \ion{He}{i} lines (see Sect.~\ref{sec:HeI}), and the [\feii{}] forest (from the N3 emission complex in I Zw1; \citealt{veron-cetty2004}).

\section{Balmer lines fitting}

\begin{figure}
    \centering
    \includegraphics[width=\linewidth]{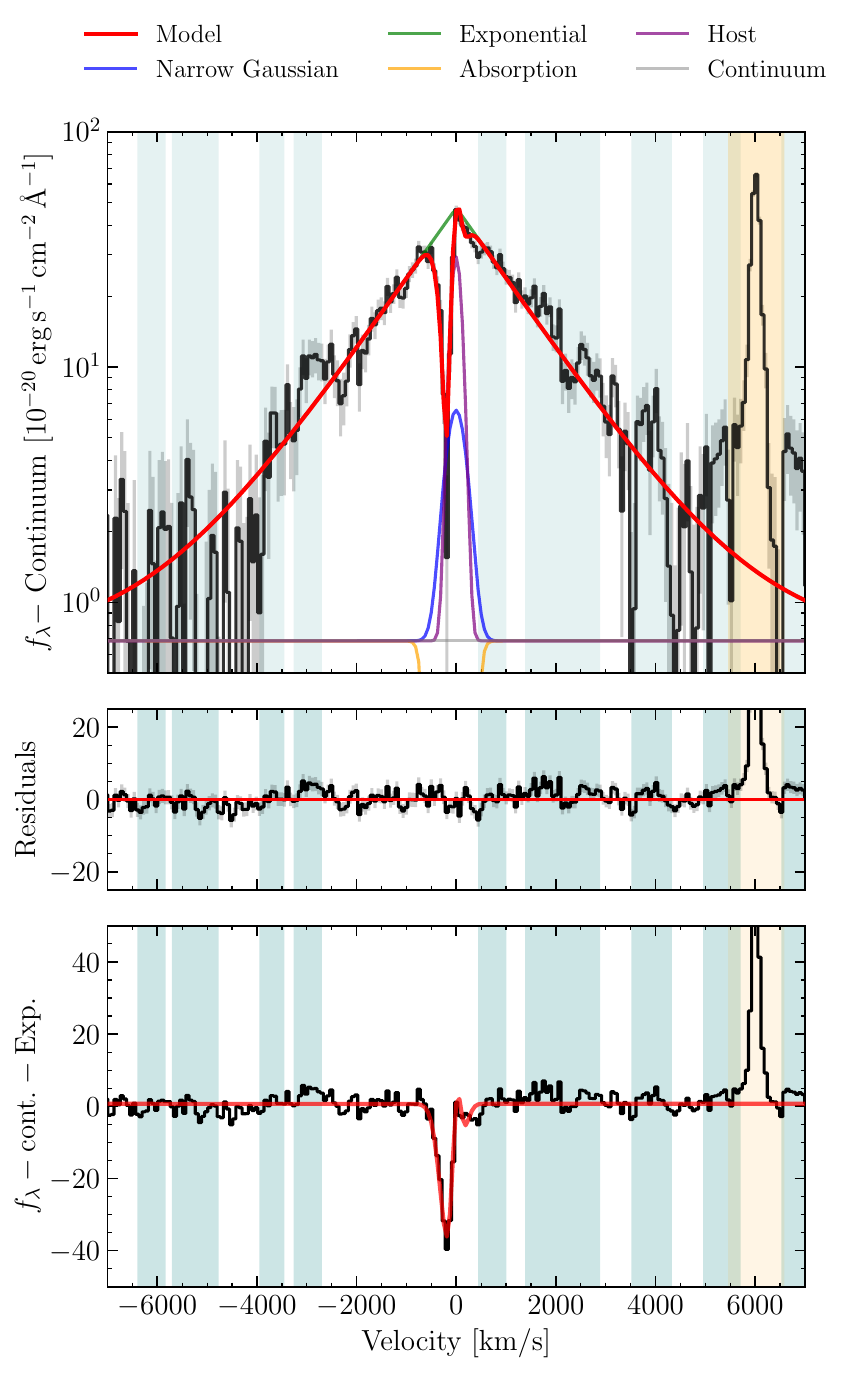}
    \caption{Same as Fig.~\ref{fig:Hb_line_fit}, but in this case we fixed the exponential scale parameter of the broad component to be the same as in the best fit of \Halpha{} ($\rm BIC = 68$).}
    \label{fig:Hb_fit_fixed_exp}
\end{figure}

In Fig.~\ref{fig:Hb_fit_fixed_exp} we show the best-fit for the \Hbeta{} model described in Sect.~\ref{sec:balmer_O3}, but here we fix the exponential scale of the broad component to the one of the H$\alpha$ line. As discussed in the formerly mentioned section, by fixing the scale we obtain a fit which is statistically indistinguishable from the best-fit letting this parameter free.

\begin{figure}
    \centering
    \includegraphics[width=\linewidth]{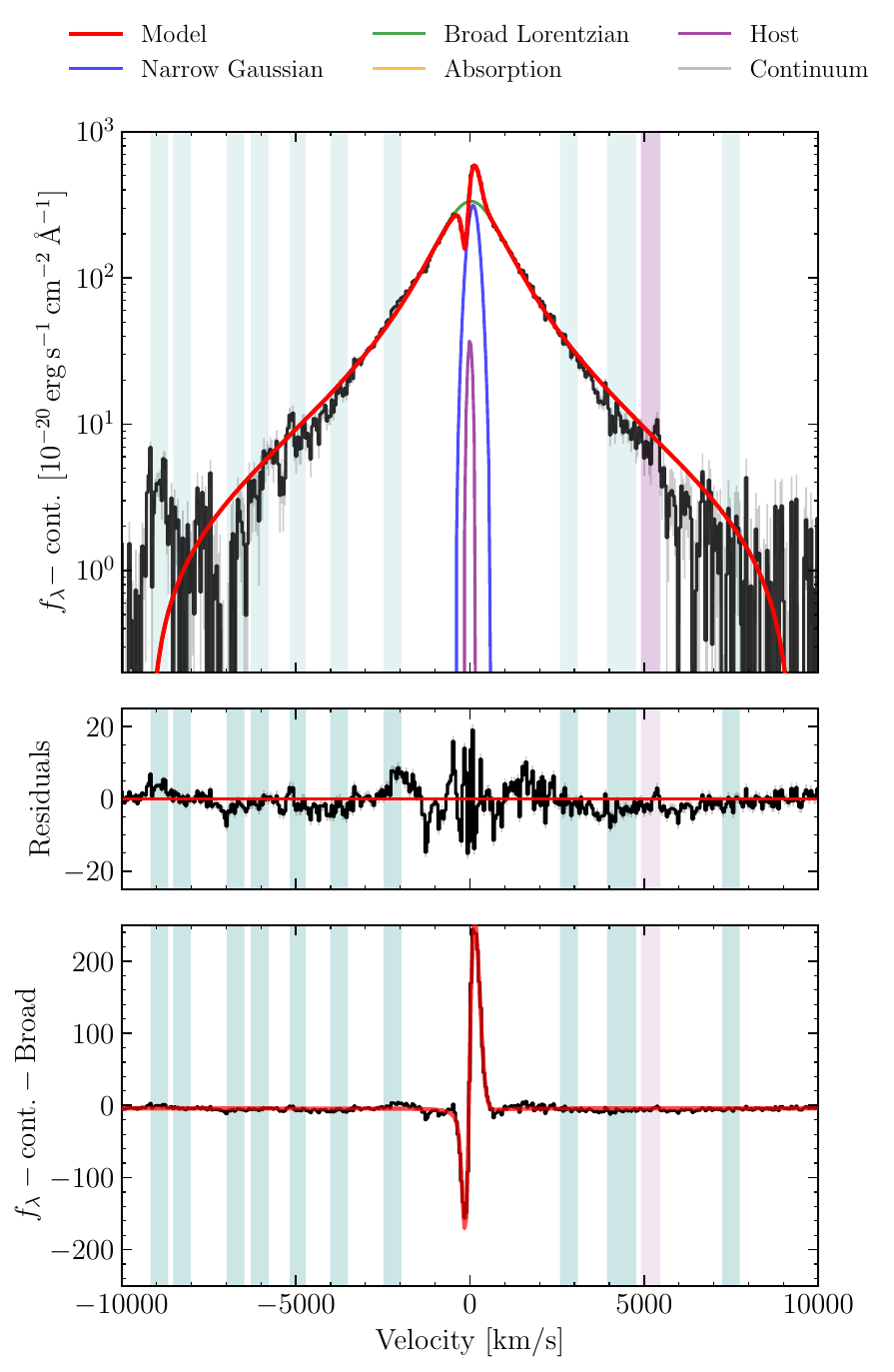}
    \caption{Same as Fig.~\ref{fig:Ha_line_fit}, but replacing the exponential component with a broad Lorentzian ($\rm BIC = 1027$).}
    \label{fig:Ha_line_fit_lorentz}
\end{figure}

\begin{figure}
    \centering
    \includegraphics[width=\linewidth]{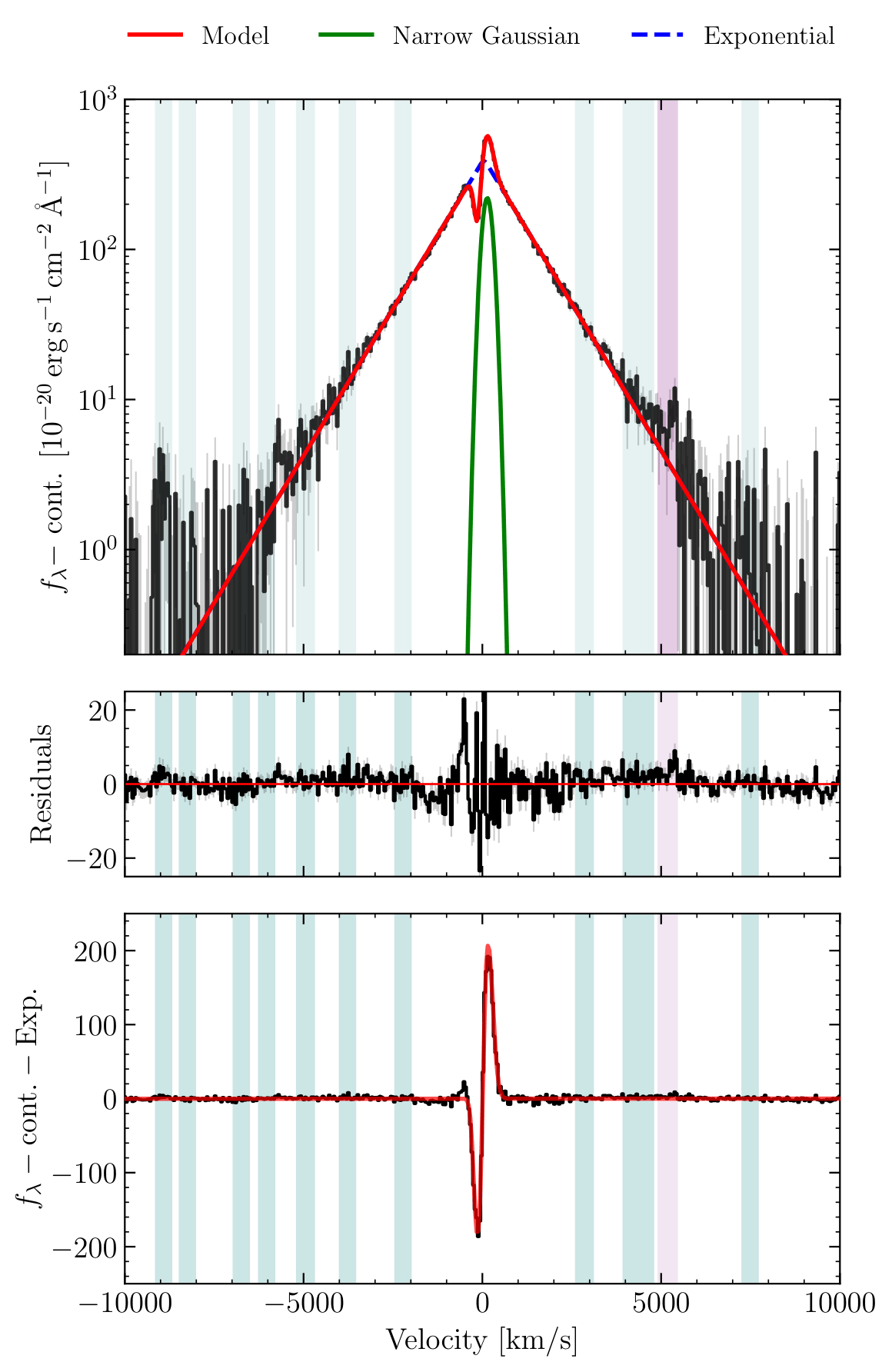}
    \caption{Same as Fig.~\ref{fig:Ha_line_fit}, but setting the amplitude of the host Gaussian component to zero ($\rm BIC = 931$).}
    \label{fig:Ha_line_fit_no_host}
\end{figure}

\begin{figure}
    \centering
    \includegraphics[width=\linewidth]{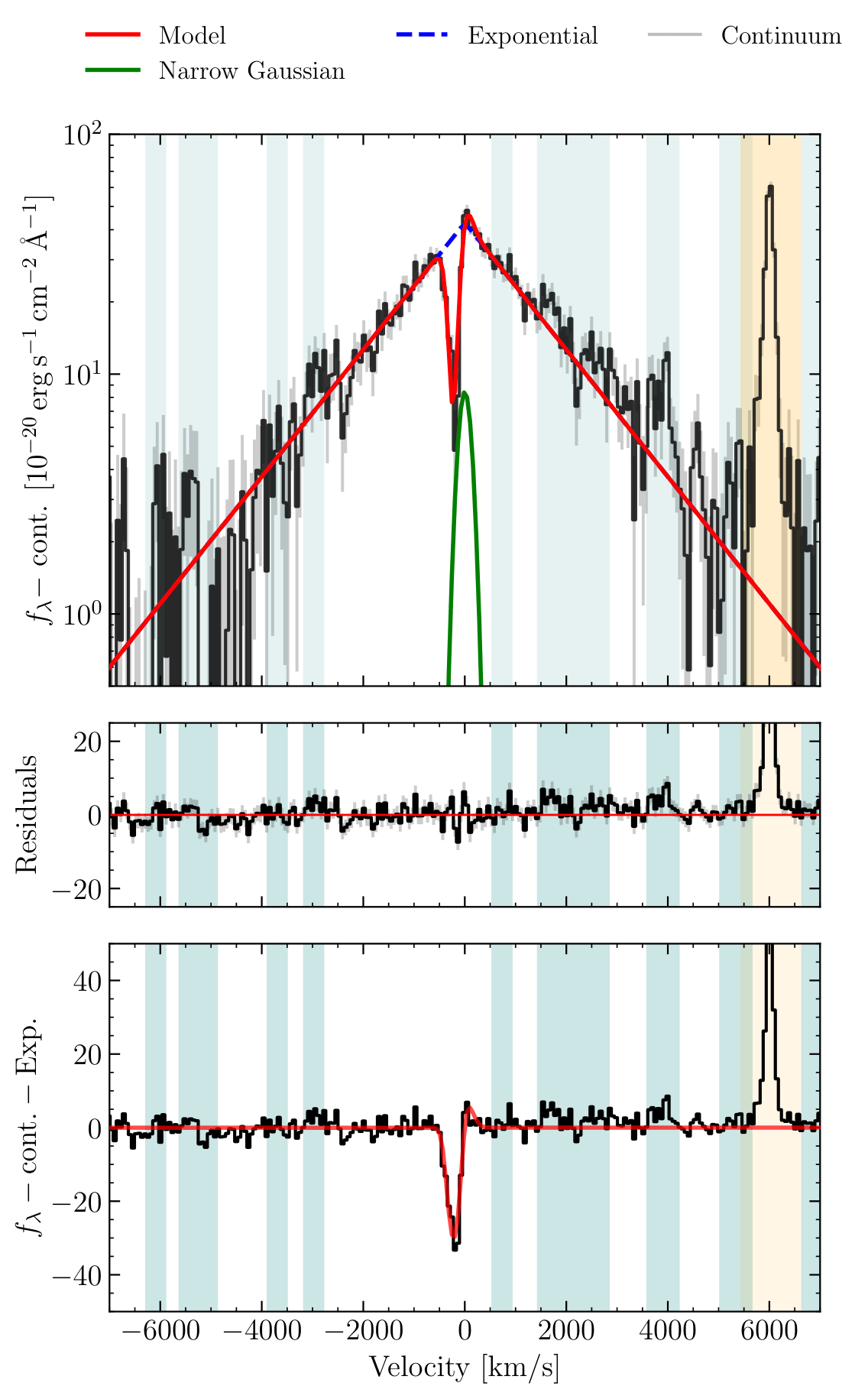}
    \caption{Same as Fig.~\ref{fig:Hb_line_fit}, but setting the amplitude of the host Gaussian component to zero ($\rm BIC = 68$).}
    \label{fig:Hb_line_fit_no_host}
\end{figure}

\section{[\feii{}] fit 6350--8000~\AA{}}

\begin{figure*}
    \centering
    \includegraphics[width=\linewidth]{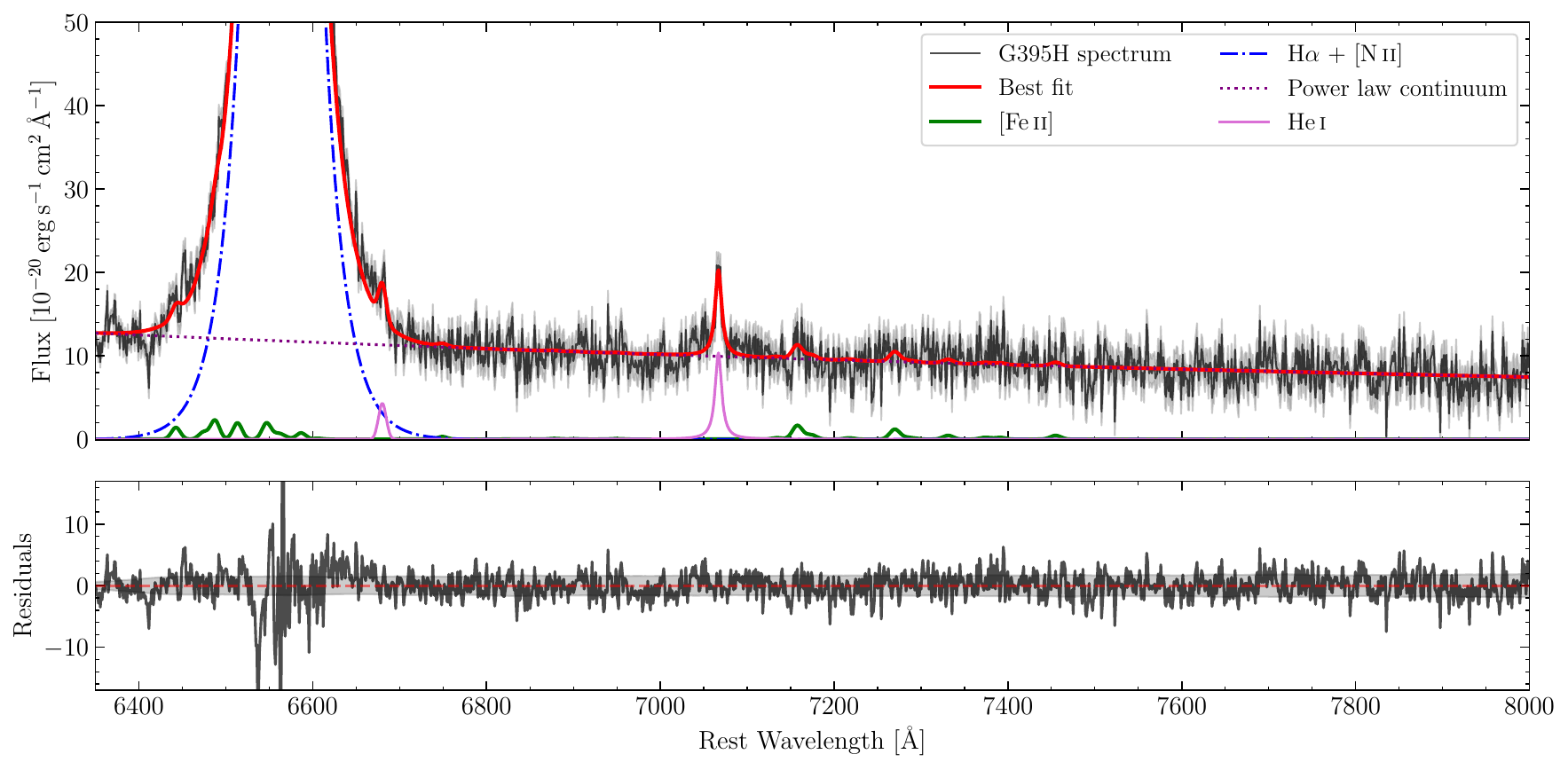}
    \caption{Zoomed-in image of the G395H spectrum of GN-9771, in the 6350--8000~\AA{} range, containing \Halpha{} and \ion{He}{i} $\lambda$7067, among other lines. As in Fig.~\ref{fig:feii_fit_1}, we show the best fit of [\feii{}] (green line).}
    \label{fig:feii_fit_ha}
\end{figure*}

In Fig.~\ref{fig:feii_fit_ha} we show the [\feii{}] fit as described in Sect.~\ref{sec:forb_feii_model}.
The fluxes of the fit [\feii{}] lines are weaker than in the 4400--6000~\AA{} range, moreover the strongest fit lines are blended with \Halpha{}.

\section{BH* temperature and electron density}

\begin{figure}
    \centering
    \includegraphics[width=\linewidth]{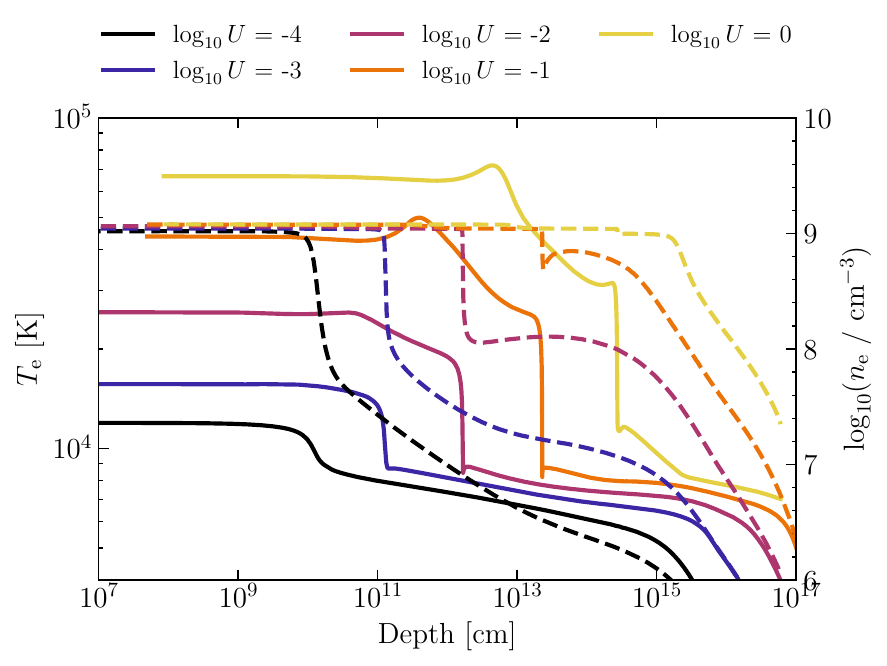}
    \caption{{\bf Temperature-density structure of a suite of BH* models, for varying ionization parameter.} Same as Fig.~\ref{fig:BHS_Te_ne}, but for a fixed $\Hden=10^9$~\unit{cm^{-3}}, for different choices of the ionization parameter \logU{}. The solid lines indicate temperature (left axis) and the dashed lines indicate electron density (right axis). The ionization parameter mainly influences the temperature of the gas before the ionization front. All the models present a thermalized outer layer with $\Te \approx 6000$--7500~K, independently of \logU{}.}
    \label{fig:BHS_Te_logU}
\end{figure}

In Fig.~\ref{fig:BHS_Te_logU} we show the \Te{} and \eden{} profiles for photoionized clouds with $\Hden = 10^{10}$~\unit{cm^{-3}}, for different choices of the ionization parameter \logU{}. All models display qualitatively similar behaviour with a thermalized outer layer with a temperature $\Te \approx 6000$--7500~K.

\end{document}